\documentclass[aps,pra,twocolumn,showpacs,amsmath,amssymb,superscriptaddress]{revtex4-1}

\usepackage{amsmath,amssymb}
\usepackage{graphicx}
\usepackage{dsfont}
\usepackage{mathrsfs}
\usepackage{dcolumn}
\usepackage{bm}
\usepackage{ulem} 

\usepackage{subfigure}
\usepackage{amsthm}
\usepackage{hyperref}
\usepackage{xcolor}
\usepackage{bbold}
\usepackage{float}

\newcommand{\x}{\mathbf{r}}

\newcommand{\tY}[2]{\tilde Y_{#1,\,#2}}

\newcommand{\revM}[1]{#1}
\newcommand{\hia}[1]{}

\hypersetup{colorlinks=true, linkcolor=black, citecolor=black, urlcolor=blue}
\allowdisplaybreaks

\begin{document}

\title{{Quantum} reactive scattering in the long-range ion-dipole potential}
\author{Tomasz Wasak}
\affiliation{Max-Planck-Institut f\"ur Physik komplexer Systeme, N\"othnitzer~Str.~38, 01187 Dresden, Germany}
\affiliation{Faculty of Physics, University of Warsaw, ul. Pasteura 5, PL-02-093 Warszawa, Poland}
\author{Zbigniew Idziaszek}
\affiliation{Faculty of Physics, University of Warsaw, ul. Pasteura 5, PL-02-093 Warszawa, Poland}

\begin{abstract}
An ion and a polar molecule interact by an anisotropic ion-dipole potential scaling as $- \alpha \cos (\theta)/r^2$ at large distances. Due to its long-range character, it modifies the properties of angular wave functions, which are no longer given by spherical harmonics. In addition, an
effective centrifugal potential in the radial equation can become attractive for low angular momenta. In this paper, we develop a general framework for an ion-dipole
reactive scattering, focusing on the regime of large~$\alpha$. We introduce modified spherical harmonics as solutions of the angular part of the Schr\"odinger equation and
derive several useful approximations in the limit of large $\alpha$. We present a formula for the scattering amplitude expressed in terms of the
modified spherical harmonics and we
derive expressions for the elastic and reactive collision rates. The solutions of the radial equation are given by Bessel functions, and we analyse their behaviour in two
distinct regimes corresponding, basically, to attractive and repulsive long-range centrifugal potentials. Finally, we study reactive collisions in the universal regime,
where the short-range probability of loss or reaction is equal to unity.
\end{abstract}

\maketitle
\section{Introduction}

Hybrid systems involving cold atoms and ions are gaining increasing attention both in theory and experiment~\cite{Tomza2019}. On one hand, recent experiments have
succeeded in combining ions confined in radio-frequency traps with ultracold atomic gases stored in optical
potentials~\cite{Smith2005,Grier2009,Zipkes2010a,Harter2010,Hall2011,Sullivan2012,Ravi2012,Feldker2020}, or producing charged particles directly in the ultracold gas via
Rydberg excitations~\cite{Kleinbach2018}. On the other hand, theoretical proposals have shown the relevance of such systems for a number of applications, ranging from
implementation of quantum gates~\cite{Doerk2010atom,Nguyen2012,Secker2016} and quantum simulations~\cite{Bissbort2013,Gerritsma2012,Joger2014}, realization of new
mesoscopic quantum states \cite{Cote2002,Massignan2005}, probing quantum gases~\cite{Sherkunov2009,Goold2010,Schurer2014,Schurer2015} to fundamental studies of low-energy
collisions and molecular
states~\cite{Idziaszek2007controlled,Idziaszek2011a,Gao2010,Gao2011,Gao2013,Simoni2011,Melezhik2016,Melezhik2019,Krych2011,Tomza2015,Tomza2015Cr,Gacesa2017}. Much recent
work has been focused on studying controlled chemical reactions at low temperatures in such systems~\cite{Rellegert2011,Hall2011,Hall2012,Felix2013,Joger2017}.

Another powerful platform for fundamental research in quantum physics are ultracold gases of molecules \cite{Carr2009,Quemener2012}. Trapping of ultracold polar molecules
in optical lattices leads to a variety of novel quantum phases or can be applied to perform quantum computations \cite{Carr2009,Quemener2012}. So far, the quantum
degenerate regime has been achieved only for bialkali dimers \cite{Lang2008,Ni2008,Danzl2010,Molony2014,Takehoshi2014,Park2015,Guo2016,Rvachov2017}. Bialkali molecules in
the ro-vibrational ground state can be classified into reactive and non-reactive ones \cite{Zuchowski2010}. While reactive collisions can be explained by relatively
simple quantum scattering models based on the properties of the long-range potential~\cite{Idziaszek2010,Idziaszek2010a}, collisions of nonreactive molecules are far more
complicated, as the scattering is affected by the presence of a dense spectrum of overlapping resonances, leading to the so-called sticky
collisions~\cite{Gregory2019}. Their theoretical treatment is based on methods derived from random-matrix theory \cite{Mayle2012,Mayle2013}.  Ultracold chemical reactions
of molecules can be controlled by external fields \cite{Ni2010}, internal spin states \cite{Ospelkaus2010}, or by aligning them in optical lattice structures of reduced
dimensions \cite{Micheli2010,Quemener2010,Quemener2011,Julienne2011,Zhu2013,Simoni2015}.  So far they have been studied experimentally in KRb
\cite{Ospelkaus2010,Ni2010,DeMarco2019}, NaLi \cite{Rvachov2017} and triplet Rb$_2$ \cite{Drews2017}.  Modern techniques in manipulation of single atoms in optical
tweezers, have allowed to assembly ultracold molecules directly from two atoms in a single, controlled chemical reaction \cite{Liu2018}.

Recently, first steps have been done towards combining cold polar molecules with cold molecular ions in a single experimental setup \cite{Eberle2015,Haas2019}. Motivated by
these attempts, in this work we study quantum scattering of an ion with a polar molecule in the low-energy regime. Here, we consider only collisions in the long-range
part of the interaction given by the ion-dipole potential and assume fixed orientation of the electric dipole moment in the course of the collision. We note, that in
general spatial orientation of a molecule varies in time, and even ultracold polar molecules in the ground state of rotational motion rotate having no net electric dipole
moment.  In this sense, our study is a necessary prerequisite before performing more elaborated analysis including rotational degrees of freedom, and the effects of
molecule polarization by the ion's charge or an external electric field. Hence, the full description of the scattering problem would require solving a set of
close-coupled equations expanded in the basis of rotational states of a molecule. In this context, our solutions derived in the current paper can be useful as an expansion
basis of the relative motion, describing long-range behaviour of the wave function components.

Another situation when our treatment is directly applicable is the collision of a very light charged particle, like an electron or a positron, with a heavy molecule. In
such collisions the electron (positron) energy is much higher than the rotational constant, and the scattering calculations can be done for a fixed orientation of the
polar molecule.  For such systems, ion-dipole collisions have been systematically studied, in particular, in the low-energy
regime~\cite{Altshuler1957,Mittleman1965,Crawford1967,Fabrikant1976,Lane1980,Garrett1981,Fabrikant1983}. In this context, it is known that the long-range ion-dipole
potential $- \alpha \cos (\theta)/r^2$ modifies the properties of the angular momentum wave functions introducing corrections to the centrifugal potential in the radial
equation~\cite{Sadeghpour2000}. It was shown that for dipole moments larger than the critical value~$\alpha_\mathrm{cr} =1.279$, the potential becomes too attractive (at
least in some directions)~\cite{mott_massey}, and the collapse to the center takes place~\cite{Mittleman1965}.

In collisions of atomic or molecular ions and polar molecules, typical values of the parameter $\alpha$ are very large, and such a regime requires a separate analysis. So
far, ion-molecule collisions have been studied by means of a classical dynamics, semi-classical approximations or variational
methods~\cite{Dugan1973,Chesnavich1980,BudenHolzer1982,Babcock1983,Liu1991}. In this paper, we study the scattering problem for the ion-dipole potential focusing on the
regime of very large $\alpha$. In such a case, the wave functions at $r \to 0$ is singular, and one needs to impose some supplemental boundary conditions, defining the
short-range behaviour of the wave function. This could be done, for instance, in the spirit of the quantum-defect theory
(QDT)~\cite{Seaton1983,Greene1979,Greene1982,Mies1984a,Mies1984b}, where one introduces some short-range parameters, that weakly depend both on the collision energy and
on the angular momentum of the relative motion~\cite{Idziaszek2009a}.

In general, such a treatment can be extended to the case of the reactive scattering, where apart from the phase parameters, one additionally introduces amplitude of the
short-range reaction processes~\cite{Idziaszek2010,Idziaszek2010a}. For realistic collisions, the short-range QDT parameters depend on the details of the short-range
potential of the specific system. In this paper, we perform the analysis of the reactive scattering in the universal regime, when the reaction probability is equal to unity at
short range. In this very special case, there is no outgoing probability flux at small distances, and the phase of the short-range wave function is not important,
hence, there is no need to include any additional QDT parameters.

\revM{
It is worth to emphasize that due to the $1/r^2$ dependence, the ion-dipole potential exhibits a very peculiar features. Firstly, for such a potential one cannot define
any kind of characteristic length scale or the energy scale, as can be done for other power-law potentials \cite{Jachymski2013}. Therefore, the only characteristic
parameter, that can be associated with this potential, is a dimensionless $\alpha$ parameter. Secondly, the local de Broglie wavelength is $\lambda(r)/(2\pi) \sim
r/\sqrt{\alpha}$, and, therefore, for large $\alpha$ the condition for the quasi-classical approximation, i.e., $\lambda'(r)/(2 \pi) \ll 1$ \cite{landau1958course}, is fulfilled at
all distances. In this case, there is no quantum reflection process at the intermediate distances, as happens for most power-law potentials at low energies
\cite{Jachymski2013}.  Hence, the relative amplitude of the incoming and outgoing flux will be the same at short and large distances for all collision energies. This
means that for $\alpha \gg 1$, reactive scattering can be very accurately described in the quasi-classical approximation. By summing all the contributions from different
partial waves, it turns out that in the limit of large $\alpha$ the total reactive cross section is identical to the cross section calculated in the framework of the classical physics.
}

The paper is organized as follows. In Section~\ref{Sec:Separ} we show how the Schr\"odinger equation separates into the radial and angular parts. In
Section~\ref{Sec:Resol} we derive some useful properties of modified spherical harmonics, which are later used in Sec.~\ref{Sec:Scatt} to calculate the scattering
amplitude. The general formulas for elastic and reactive collision rates are derived in Sec.~\ref{Sec:Elast} and Sec.~\ref{Sec:React}, respectively. In
Sec.~\ref{Sec:SolAng} we investigate properties of modified spherical harmonics, while Sec.~\ref{Sec:SolRad} is devoted to analysis of the radial solutions. Reactive
scattering in the universal limit is discussed in Sec.~\ref{Sec:Univ}. We conclude in Sec.~\ref{Sec:Concl} presenting some final remarks. In Appendices we present
technical details of the calculations.

\section{Separation of the Schr\"odinger equation}
\label{Sec:Separ}

We consider the scattering of a molecule with a permanent electric dipole moment, and a charged particle, which could be for instance a monoatomic ion as well as an electron
or a positron. We assume that the dipole moment orientation is fixed in space, which is equivalent to solving the equations of motion in a body-fixed frame related to the
polar molecule. The Schr\"odinger equation describing the wave function of the relative motion in the ion-dipole potential reads
\begin{equation}
\label{schroed}
	\frac{\hbar^2}{2 \mu} \Big( - \frac{1}{r^2}\frac{\partial}{\partial r} r^2 \frac{\partial}{\partial r}  + \frac{\hat l^2}{r^2} \Big) \psi(\x) + V_\mathrm{id}(\x) \psi(\x) = E \psi(\x),
\end{equation}
where $\mu$ is the reduced mass of the particles, $V_\mathrm{id}$ is the interaction potential between the ion with a charge $q$ and a polar molecule with a dipole moment $\mathbf d$.
The square of the angular momentum operator is given by 
\begin{equation}
\hat l^2 = 
	- \frac{1}{\sin\theta} \frac{\partial}{\partial \theta} \bigg(\!\! \sin\theta \frac{\partial}{\partial \theta} 
	\bigg) - \frac{1}{\sin^2\theta} \frac{\partial^2}{\partial \phi^2}.
\end{equation}
The interaction potential is given by the scalar product of the dipole moment $\mathbf d$ and the electric field $\mathbf E$ of the ion:
\begin{equation}
	V_\mathrm{id}(\x) = - \mathbf{d}\cdot \mathbf{E} = - \frac{q \mathbf{d} \cdot \x}{4\pi\epsilon_0 r^3}.
\end{equation}
Denoting the angle between the vectors $\mathbf{d}$ and $\mathbf{E}$ by $\theta$, we have
\begin{equation}
	V_\mathrm{id}(\x) = - \frac{q d \cos\theta}{4\pi\epsilon_0 r^2}.
\end{equation}
Introducing $E = \hbar^2 k^2 / 2 \mu$, the  Schr\"odinger equation can be rewritten as
\begin{equation}
\label{seq1}
	 \Big( - \frac{1}{r^2}\frac{\partial}{\partial r} r^2 \frac{\partial}{\partial r}  + \frac{\hat l^2}{r^2} - \frac{\alpha \cos\theta}{r^2} \Big) \psi(\x) = k^2 \psi(\x),
\end{equation}
where we defined the dimensionless parameter $\alpha$ as
\begin{equation}
\label{alpha}
	\alpha \equiv \frac{qd}{4\pi\epsilon_0}\frac{2 \mu}{\hbar^2}.
\end{equation}

In physical systems composed of a polar molecule and a monoatomic ion, a permanent dipole moment is of the order of $1 D$ (Debye). For such systems, $\alpha$ is usually a large number. From the point of view of recent experiment on ultracold systems, the most relevant are polar molecules of two-alkali metal atoms and alkaline earth metal ions. In such a case  
$\alpha$ ranges from $\alpha = 4.43 \times 10^3$ for LiNa--$^9$Be$^+$ up to $\alpha = 6.07\times 10^5$
for LiCs--$^{174}$Yb$^+$. In contrast, for electron/positron scattering on a molecule with a permanent dipole moment, $\alpha$ is typically of the order of one.

The system possess the cylindrical symmetry, so the quantum number $m$ associated with the $z$-component of the angular momentum is conserved.
In general the wave function $\psi(\x)$ can be decomposed into a radial $R(r)$ and an angular $\tilde Y_m(\theta,\phi)$ parts
\begin{equation}
	\psi(\x) = R_m(r) \tilde Y_m(\theta,\phi).
\end{equation}

The centrifugal barrier and the dipole-ion potential fall off with the distance according to the same power law. Therefore the radial and angular part of the wave
functions can be solved independently. We introduce the operator
\begin{equation}
	\hat U \equiv \hat l^2 - \alpha \cos\theta
\end{equation}
\revM{that describes the angular part of the stationary states $\tilde Y_{\ell,m}(\theta,\phi)$, where $\ell$ numbers the eigenvalues of $\hat U$}. The solution of the eigenvalue problem
\begin{equation}
\label{angular}
	(\hat l^2 - \alpha \cos\theta)\tilde Y_{\ell,\,m}(\theta,\phi) = \lambda_{\ell,m} \tilde Y_{\ell,m}(\theta,\phi),
\end{equation}
gives the spectrum $\lambda_{\ell,m}$ and the corresponding eigenfunctions $\tilde Y_{\ell,m}(\theta,\phi)$, which in the rest of the paper will be referred to as the
{\it modified spherical harmonics}. \revM{We choose the numbering of the eigenvalues as $\ell=|m|, |m|+1,|m|+2, \ldots$, to recover the standard spherical harmonics $Y_{\ell,m}(\theta,\phi)$ from 
$\tilde Y_{\ell,m}(\theta,\phi)$ in the limit of vanishing $\alpha$.}
Similarly to the standard spherical harmonics, we \revM{impose the normalization condition}
\begin{equation}
\int d \Omega \,\tilde Y^*_{\ell,m}(\theta,\phi) \tilde Y_{\ell',m'}(\theta,\phi) = \delta_{\ell',\ell} \delta_{m,m'}.
\end{equation}

The next step is to solve the radial part of the Schr\"odinger equation given by
\begin{equation}
\label{radial}
	 \Big( - \frac{1}{r^2}\frac{\partial}{\partial r} r^2 \frac{\partial}{\partial r}  + \frac{\lambda_{\ell,m}}{r^2} \Big) R_{\ell,m}(r) = k^2 R_{\ell,m}(r).
\end{equation}
At sufficiently large distances, when the short-range potential can be entirely neglected and only the dipole-ion interaction is present, the radial part can be expressed
in terms of spherical Bessel functions of the order given by a real or purely imaginary number.

\revM{Below we investigate the angular and radial solutions. But first, we reconsider the scattering problem in terms of the modified spherical harmonics.}

\section{Resolution of plane wave in modified spherical harmonics}
\label{Sec:Resol}
Before we analyse the scattering problem, we present two formulas that are used in derivation of the scattering amplitude. The first is the resolution of the \revM{angular} identity operator:
\begin{eqnarray}
\label{identity}
	\sum_{\ell=0}^{\infty} \sum_{m=-\ell}^{\ell} 
	Y^*_{\ell,m}(\mathbf{n}_1) & &Y_{\ell,m}(\mathbf{n}_2) =  
	\sum_{\ell=0}^{\infty} \sum_{m=-\ell}^{\ell} \tilde Y^*_{\ell,m}(\mathbf{n}_1) \tilde Y_{\ell,m}(\mathbf{n}_2) \nonumber\\
	& & = \frac{1}{\sin \theta_1} \delta(\theta_1-\theta_2) \delta(\phi_1-\phi_2),
\end{eqnarray}
where $\mathbf{n}_i$ are unit vectors along $\theta_i$ and $\phi_i$ ($i=1,2$). The vectors $\mathbf{n}_1$ and $\mathbf{n}_2$ can be interchanged on each side without affecting the sums.
Also, the position of the complex conjugate is unimportant. This formula represents the fact that eigenvectors of operator $\hat U$ form a complete orthogonal basis (for fixed value of $\alpha$).

The second formula is the expansion of the plane wave~$e^{i \mathbf{k}.\mathbf{r}}$ in the basis of the modified spherical harmonics $\tilde Y_{\ell,m}$.
We start from the familiar expansion (see e.g.~\cite{mott_massey})
\begin{equation}
	e^{i \mathbf{k}{.}\mathbf{r}}  = 4\pi \sum_{\ell=0}^{\infty} \sum_{m=-\ell}^{\ell} i^\ell j_\ell(kr) Y^*_{\ell,m}(\hat{\mathbf{k}})Y_{\ell,m}(\hat{\mathbf{r}}),
\end{equation}
where $j_\ell(x)$ is the spherical Bessel function of the first kind, $\hat{\mathbf{k}}$ and $\hat{\mathbf{r}}$ are unit vectors (denoted by hat) directed along
${\mathbf{k}}$ and ${\mathbf{r}}$, respectively.  For large values of $r = |\mathbf{r}|$ this expansion takes the following form:
\begin{equation}
	e^{i \mathbf{k}.\mathbf{r}}  \xrightarrow[kr \gg 1]{} 4\pi \sum_{\ell,m}  \frac{\sin\big(kr - \frac{\ell \pi}{2}\big)}{kr} i^l Y^*_{\ell,m}(\hat{\mathbf{k}})Y_{\ell,m}(\hat{\mathbf{r}}).
\end{equation}
This can be rewritten as
\revM{
\begin{eqnarray}
\label{plane_wave_Y}
	&&e^{i \mathbf{k}.\mathbf{r}} \xrightarrow[kr \gg 1]{} \frac{4\pi}{2ikr} \sum_{\ell,m} \big(e^{ikr} \! - \! (-1)^\ell e^{-ikr}\big) Y^*_{\ell,m}(\hat{\mathbf{k}})Y_{\ell,m}(\hat{\mathbf{r}}) \nonumber \\
	&& = \frac{4\pi}{2ikr}\! \sum_{\ell,m}\! \bigg(\! e^{ikr} Y^*_{\ell,m}(\hat{\mathbf{k}})Y_{\ell,m}(\hat{\mathbf{r}}) \! - \! e^{-ikr}  Y^*_{\ell,m}(-\hat{\mathbf{k}})Y_{\ell,m}(\hat{\mathbf{r}})\! \bigg) \nonumber.\\
	& & 
\end{eqnarray}
}
Here, we have used the parity of the spherical harmonics $(-1)^\ell Y_{\ell,m}(\hat{\mathbf{k}}) = Y_{\ell,m}(-\hat{\mathbf{k}})$.
Employing  Eq.~(\ref{identity}) in the second line in the above formula, we arrive at 
\revM{

\begin{equation}
\begin{split}
\label{plane_wave_tY}
	e^{i \mathbf{k}\cdot\mathbf{r}} \xrightarrow[kr \gg 1]{} &\\
	\frac{2\pi}{ikr}\! \sum_{\ell,m}\!\! \bigg(\!\!\, &
		 e^{ikr} \tilde{Y}^*_{\ell,m}(\hat{\mathbf{k}})\tilde{Y}_{\ell,m}(\hat{\mathbf{r}})
			-  e^{-ikr}  \tilde{Y}^*_{\ell,m}(\!-\hat{\mathbf{k}})\tilde{Y}_{\ell,m}(\hat{\mathbf{r}})
			\!\! \bigg). 
\end{split}
\end{equation}
}
In the second term, the minus sign in front of $\hat{\mathbf{k}}$ could be equivalently put in front of $\hat{\mathbf{r}}$. This formula cannot be further
simplified, because modified spherical harmonics $\tilde{Y}_{\ell,m}(\hat{\mathbf{r}})$ in general do not have specified parity.

\section{Scattering problem}
\label{Sec:Scatt}
In the scattering problem we solve the Schr\"odinger equation \eqref{schroed} with the boundary conditions
\begin{equation}
\label{scatt_prob}
 \psi(\mathbf{r})  \xrightarrow[|r|\ \to \infty]{} e^{i \mathbf{k}_i \cdot \mathbf{r} } + f(\mathbf{k}_i \to \mathbf{k}) \frac{e^{i k r}}{r},
\end{equation}
where $\mathbf{k}_i$ is the initial wave vector of the incident particle, $\mathbf{k} = k \hat{\mathbf{r}}$ and $f(\mathbf{k}_i \to \mathbf{k})$ is the scattering amplitude describing the scattering process.
Note,  that the above wave function is normalized such that the probability current of the incident particle (plane wave $e^{i\mathbf{k}_i\cdot\mathbf{r}}$) is $\hbar \mathbf{k}_i / \mu$,
which is its velocity.

Below, we express the amplitude $f$ in terms of the modified spherical harmonics $\tilde{Y}_{\ell,m}$ and the elements $S_{\ell,m}$ of the scattering $S$-matrix defined as follows.
After solving the radial part of the Schr\"odinger equation~\eqref{radial}, we find its asymptotic form for large $r$:
\begin{equation}
\label{radial_R}
	R_{\ell,m}(r)   \xrightarrow[|r|\ \to \infty]{} \frac{i}{2kr}\bigg( e^{-i \big(kr - \frac{\ell \pi}{2}\big)} - S_{\ell,m} e^{i \big(kr - \frac{\ell \pi}{2}\big)} \bigg) A_{\ell,m},
\end{equation}
where $A_{\ell,m}$ is given by the boundary conditions of the considered physical problem, and $S_{\ell,m}$ are the elements of the scattering $S$-matrix.

\revM{To find the scattering amplitude, we write the wavefunction of the particles as follows:}
\begin{equation}
	\psi(\mathbf{r}) = \sum_{\ell,m} R_{\ell,m}(r) i^\ell \tilde{Y}_{\ell,m}(\hat{\mathbf{r}}),
\end{equation}
where $R_{\ell,m}(r)$ is given by \eqref{radial_R}. Setting the coefficients $A_{\ell,m} = 4\pi (-1)^\ell\tilde{Y}_{\ell,m}^*(-\hat{\mathbf{k}}_i)$ \revM{and employing Eq.~\eqref{plane_wave_tY}},
we find the relation \eqref{scatt_prob},
where the scattering amplitude is given by
\begin{equation}
\label{scatt_amp}
	f(\mathbf{k}_i \!\!\to \!\mathbf{k})\! =\!
		\!\frac{4\pi}{2ik} \! \sum_{\ell,m} \! \bigg(
			e^{i\pi\ell} S_{\ell,m} \tilde{Y}_{\ell,m}^*(-\hat{\mathbf{k}}_i)
			-  \tilde{Y}_{\ell,m}^*(\hat{\mathbf{k}}_i)
		\!\bigg) \tilde{Y}_{\ell,m}(\hat{\mathbf{k}}) .
\end{equation}

In the following sections we present the formulas for the elastic and reactive collision rate constants~$K^{\mathrm{el}}$ and~$K^{\mathrm{re}}$.

\section{Elastic collision rate~$K^{\mathrm{el}}$}
\label{Sec:Elast}

The radial coordinate $j_r$ of the probability current for the term $f e^{ikr} / r$, describing the scattering wave, is given by $j_r = \hbar k |f|^2/(\mu r^2)$. 
The differential elastic cross section $d\sigma_\mathrm{el}/d\Omega$ is defined by the following relation: $d \sigma_\mathrm{el}(\mathbf{k}_i \to \mathbf{k})= (d\sigma_\mathrm{el}/d\Omega) d\Omega$, where $d\sigma_\mathrm{el}(\mathbf{k}_i \to \mathbf{k})$ is the differential part of the total cross section, that contributes to the scattering into the solid angle $d\Omega$. 
By equating the number of particles scattered into the solid angle $d \Omega$ per unit time: $j_r d\Omega r^2$, with the number of particles in the incident particle probability flux $d\sigma_\mathrm{el} v_i$, we obtain $d\sigma_\mathrm{el}(\mathbf{k}_i \to \mathbf{k})  = j_r/v_i r^2 d\Omega$,
where $v_i = \hbar k_i/\mu= \hbar k/\mu$ is the velocity of incident particles, and we assume that probability density in the incident wave function is normalized to unity: $\rho = |\psi|^2= 1$, in accordance with normalization of the wave function assumed in \eqref{scatt_prob}. 
Hence, $d\sigma_\mathrm{el}(\mathbf{k}_i \to \mathbf{k}) = |f|^2d\Omega$.
The total elastic cross section is obtained by integrating $d\sigma_\mathrm{el}(\mathbf{k}_i \to \mathbf{k})/d\Omega$ over all possible directions $\hat{\mathbf{k}}$ of the scattered particle,
$\sigma_\mathrm{el}(\mathbf{k}_i) = \int\! (d\sigma_\mathrm{el}(\mathbf{k}_i \to \mathbf{k})/d\Omega)\,d\Omega$.
Inserting here the formula \eqref{scatt_amp}, we arrive at
\begin{equation}
\label{sigma_el_tot}
	\sigma_\mathrm{el}(\mathbf{k}_i)
	=
	\!\frac{(2\pi)^2}{k^2} \! \sum_{\ell,m} \! \Big|
			e^{i\pi\ell} S_{\ell,m} \tilde{Y}_{\ell,m}^*(-\hat{\mathbf{k}}_i)
			-  \tilde{Y}_{\ell,m}^*(\hat{\mathbf{k}}_i)
		\!\Big|^2.
\end{equation}
This final expression for the total cross section depends on the direction of the incident particle $\hat{\mathbf k}_i$. It is thus natural to consider the cross section averaged over
all possible directions of incidence. Therefore, the averaged elastic cross section is given by $\sigma_\mathrm{el}(\mathbf k_i)$ averaged over direction of $\mathbf k_i$:
\begin{equation}
\label{av_sigma_el}
	\bar{\sigma}_\mathrm{el} = \frac{1}{4\pi} \int \sigma_\mathrm{el}(\mathbf k_i) d\Omega_i.
\end{equation}
Note that the $\bar{\sigma}_\mathrm{el}$ does not depend only on $S_{\ell,m}$ but also on the form of $\tilde Y_{\ell,m}$. This comes from the fact that the potential
$V_\mathrm{id}(\x)$ is not spherically symmetric. This is mathematically expressed by the fact that the scalar product $\int \tilde{Y}_{\ell,m}^*(-\hat{\mathbf{k}}_i)
\tilde{Y}_{\ell,m}(\hat{\mathbf{k}}_i) d\Omega_i$ in general depends on $\alpha$, whereas in the case of spherical harmonics we have $\int
Y_{\ell,m}^*(-\hat{\mathbf{k}}_i) Y_{\ell,m}(\hat{\mathbf{k}}_i) d\Omega_i = (-1)^\ell$.  Nevertheless, an important simplification comes from the presence of the cylindrical
symmetry. Namely, the function $\sigma_\mathrm{el}(\mathbf{k}_i)$ depends only on angle~$\theta_i$ between the dipole moment~$\mathbf{d}$ and~$\mathbf{k}_i$. This can be
seen from \eqref{sigma_el_tot}, where the phase $\phi$ enters only as a phase $e^{i m \phi_i}$ which is unimportant after taking the modulus squared. Consequently,
in Eq.~\eqref{av_sigma_el} only the integration over one variable $\theta_i$ has to be performed.

Here, we will give an expression for the elastic collision rate constant ${\cal K}^\mathrm{el}$, which is by definition given by the averaged elastic cross section and the velocity of the incident particle:
\begin{equation}
	{\cal K}^\mathrm{el} = v_i \bar{\sigma}_\mathrm{el},
\end{equation}
An alternative formulation involves the probability flux $j_r^\mathrm{scatt}$ of the scattered particle. It is given by
\begin{equation}
\label{Kel_def}
	{\cal K}^\mathrm{el} = \frac{1}{4\pi}\int\! d\Omega_i\! \int\! d\Omega\, j_r r^2,
\end{equation}
taken at the limit $r\to\infty$. Following the normalization assumed in \eqref{scatt_prob}, we have assumed that the flux of the incident particles is equal to the velocity $\mathbf{v}_i = \hbar \mathbf{k}_i / \mu$, i.e., one particle per unit area per unit time.

It can be shown that for a pure ion-dipole potential, the total scattering cross section is infinite, which is a consequence of its long-range character and the anisotropy.
Mathematically, it is related to the fact, that in the expression for the cross section, given by Eq.~\eqref{sigma_el_tot}, 
a mixed term $\tilde{Y}_{\ell,m}(-\hat{\mathbf{k}}_i) \tilde{Y}_{\ell,m}^*(\hat{\mathbf{k}}_i)$ appears. Without the anisotropy, the modified spherical harmonics 
are equal to the standard ones, and this term is exactly $(-1)^\ell$ after integration over directions of $\hat{\mathbf{k}}_i$. With the anisotropy, the term approaches the standard
value, but not sufficiently fast, to make the \revM{sum} convergent \revM{in Eq.~\eqref{sigma_el_tot}} (see the details in Appendix~\ref{app-YY}).

\section{Reactive collision rate~$\mathcal{K}^{\mathrm{re}}$}
\label{Sec:React}
The reactive rate constant $\mathcal{K}^{\mathrm{re}}$ is most easily obtained by formulating it as a lost probability flux averaged over all directions.
We rewrite Eq.~\eqref{scatt_prob} in terms of the amplitudes $f_\mathrm{out}(r)$ and $f_\mathrm{in}(r)$:
\begin{eqnarray}
  e^{i \mathbf{k}_i \cdot \mathbf{r} } + f(\mathbf{k}_i \to \mathbf{k})  \frac{e^{i k r}}{r} & &
  \xrightarrow[|r|\ \to \infty]{} \\
  & &f_\mathrm{out}(\theta,\phi) \frac{e^{ikr}}{r} + f_\mathrm{in}(\theta,\phi) \frac{e^{-ikr}}{r}. \nonumber
\end{eqnarray}
The radial component of probability flux corresponding to the outgoing and incoming particle are given by $j_r^\mathrm{\,out} = \hbar k|f_\mathrm{out}|^2/(\mu r^2)$ and $j_r^\mathrm{\,in} = \hbar k|f_\mathrm{in}|^2/(\mu r^2)$, respectively.
The difference between these currents integrated over all possible directions of the scattered particle describes the rate of the probability loss due to the reactions during the scattering process.
The reactive rate is given by that loss of the probability averaged over all directions of incidence (calculated at the limit $r\to\infty$):
\begin{equation}
\label{Kre_def}
	\mathcal{K}^\mathrm{re} = \frac{1}{4\pi}\int\! d\Omega_i\! \int\! d\Omega\, (j_r^\mathrm{\,out}-j_r^\mathrm{\,in}) r^2.
\end{equation}
It should be noted that the above formula resembles the equation~\eqref{Kel_def}.
Using the formula  for the scattering amplitude found above, see Eq.~\eqref{scatt_amp}, the reactive rate can be expressed in terms of the scattering matrix elements $S_{\ell,m}$ according to the formula:
\begin{equation}
\label{Kre_eq}
	\mathcal{K}^\mathrm{re} = \frac{\hbar k_i}{\mu} \frac{\pi}{k_i^2} \sum_{\ell,m} \Big( 1 - |S_{\ell,m}|^2 \Big).
\end{equation}
Unlike the elastic rate, this equation is similar to the usual relation for the reactive rate expressed in spherical harmonics.


\revM{
\section{Solution of the angular part}
\label{Sec:SolAng}
}

Below we present general features of the solutions for the angular part of the wave functions. The equation that we consider is expressed by
Eq.~\eqref{angular}. 
To proceed, we decompose $\hat U$ in the basis of spherical harmonics. We express $\tilde Y_{\ell,m}$ as a linear combination of spherical harmonics~$Y_{l,m}$:
\begin{equation}
\label{Ycl}
	\tilde{Y}_{\ell,m} = \sum_{\ell'=|m|}^\infty c_{\ell,\,\ell'}^{(m)} Y_{\ell',m}.
\end{equation}
\revM{Since the cylindrical symmetry is preserved, all the harmonics have the same quantum number $m$.} The coefficients $c_{\ell,\,\ell'}^{(m)}$ follow from the
diagonalization of the operator~$\hat U$.

In order to solve the eigenvalue problem, we need to evaluate the matrix elements of this operator. The operator $\hat l^2 $ is diagonal in the basis of $Y_{\ell,m}$. The matrix elements of 
the $\cos\theta$ are given by:
\begin{eqnarray}
	&&\int_0^{2\pi}\!\!\!d\phi\!\!\int_0^\pi\!\!\!d\theta\,
		Y^*_{\ell'm'}(\theta,\phi) \, \cos\theta \, Y_{\ell m}(\theta,\phi) \sin\theta  \\
		&&	\quad\quad = \delta_{m,m'} \Big( \delta_{\ell',\,\ell-1} \beta_{m,\, \ell} + \delta_{\ell',\, \ell+1} \beta_{m,\,\ell+1} \Big),
\end{eqnarray}
with $\beta_{m,\,\ell} = \sqrt{{(\ell^2-m^2)}/{(4\ell^2-1)}}$, \revM{whereas they are zero if $l\leqslant |m|$.} The eigenproblem for the matrix $\hat U$ with $\ell\geqslant |m|$ 
is expressed then by the equation:
\begin{equation}
\label{eigenprob}
	\ell'(\ell'+1) c_{\ell,\,\ell'}^{(m)} - \alpha\big( \beta_{m,\,\ell'} c_{\ell,\,\ell'-1}^{(m)} + \beta_{m,\,\ell'+1} c_{\ell,\,\ell'+1}^{(m)} \big) = \lambda_{\ell,\,m} c_{\ell,\,\ell'}^{(m)}.
\end{equation}
Below we show the results of the numerical calculations of the modified spherical harmonics $\tilde Y_{\ell,\,m}$ in the regimes $\alpha \sim 1$ and $\alpha \gg 1$. \revM{We
also present various approximations schemes for this equation in the latter limit. The details of the methods and derivations are presented in Appendix~\ref{App:SolAng}.}

\revM{
\subsection{Angular orbitals for intermediate values of $\alpha $}
In this section we investigate the properties of the angular wave functions numerically. To this end we solve Eq.~\eqref{eigenprob} by writing it in the matrix form, see also
Eq.~\eqref{matU} in Appendix~\ref{App:SolAng}, truncating the matrix dimension of $\hat U$ for final $\ell_\mathrm{max} = 4000$ and evaluate the modified spherical
harmonics for $\alpha=1$ and $\alpha=10$, assuming $m =0$.

The results are plotted in Figs.~\ref{fig:orbitals1} and \ref{fig:orbitals10}, where we display three-dimensional plots of $|\tilde Y_{\ell,m}(\theta, \phi)|^2$ as a
function of $\theta$ and $\phi$ for $\ell=0,1,2,3$. Since all the function $\tilde Y_{\ell,m}$ are proportional to $e^{i m \phi}$, the shown plots are cylidrically symmetric 
with respect to rotations about the $z$-axis.

We observe that for $\alpha=1$ the modified spherical harmonics are similar to the standard spherical harmonics, but they are slightly shifted towards the upper half-plane, which
is due to the attractive part of the ion-dipole potential. However, for $\alpha=10$ the orbitals are highly anisotropic, and they are no longer symmetric with respect to
reflections $z\to -z$. In the figure, the gray XY surface marks $z=0$ plane, and the intersection of the orbital with that plane is shown with a white contour.  In
particular, the orbitals $\ell=0$, $m=0$ for both $\alpha=1$ and $\alpha=10$ are strongly shifted to the upper half-space.

In Fig.~\ref{fig:orbital00} we present the dependence of the orbital $\ell=0$ and $m=0$ on $\alpha$. We note that for increasing value of $\alpha$, the orbitals become more anisotropic. 
Even for relatively low values $\alpha \approx 1$, the displacement of the orbital is significant compared to the isotropic case $\alpha=0$.

%
%
\begin{figure}[htb!]
\includegraphics[clip, scale=0.35]{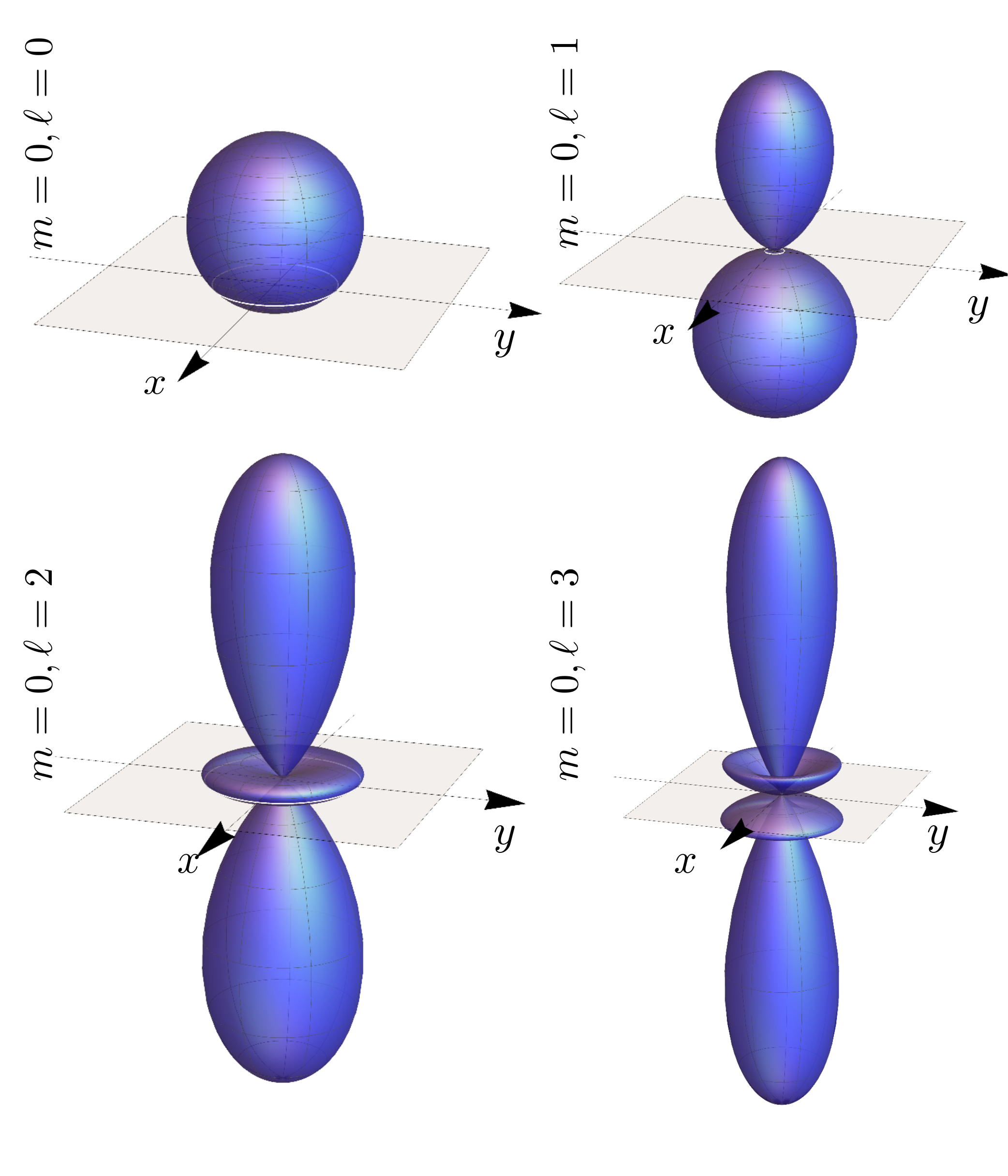}
	\caption{
		Modified spherical harmonics $|\tilde Y_{\ell,m}|^2$ as a function of spherical coordinates for $\alpha=1$, $m=0$,
		and different values of $\ell = 0,1,2,3$ (from upper left to lower right panel).
	}\label{fig:orbitals1}
\end{figure}

%
%
\begin{figure}[htb!]
\includegraphics[clip, scale=0.35]{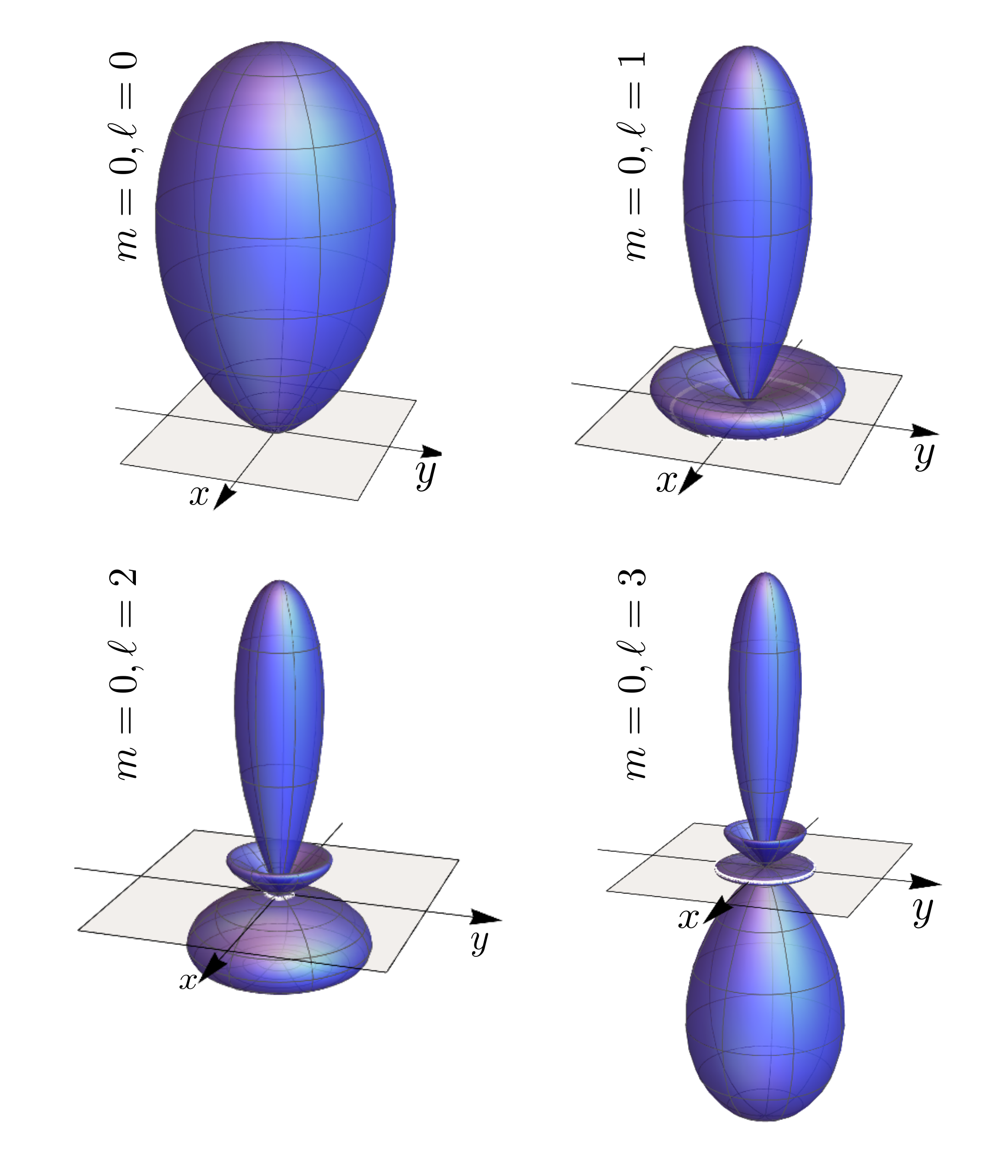}
  \caption{
    Modified spherical harmonics $|\tilde Y_{\ell,m}|^2$ as a function of spherical coordinates for $\alpha=10$, $m=0$,
    and different values of $\ell = 0,1,2,3$ (from upper left to lower right panel). 
	}\label{fig:orbitals10}
\end{figure}

%
%
\begin{figure}[htb!]
\includegraphics[clip, scale=0.35]{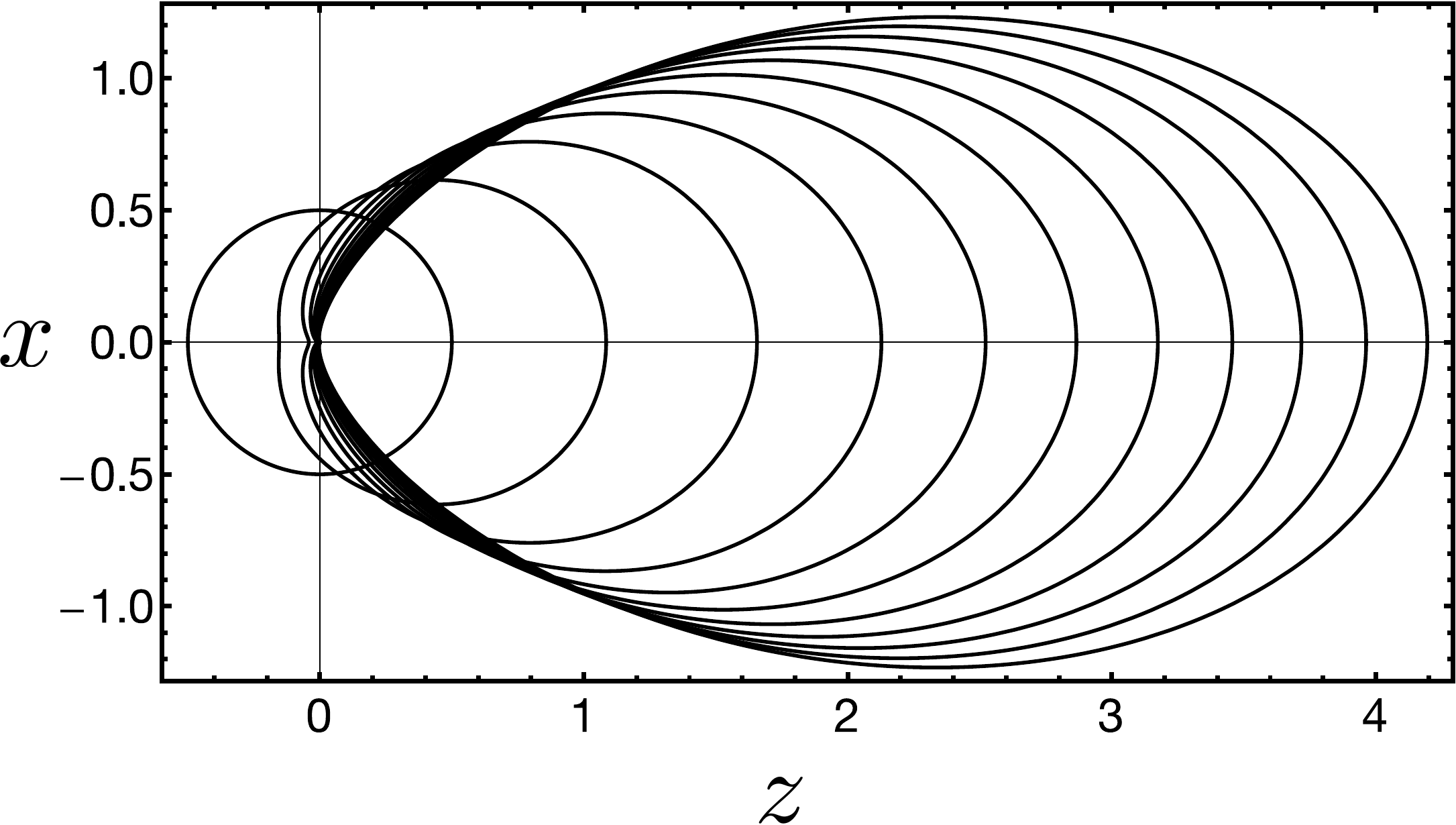}
  \caption{ The cut of the modified spherical harmonic $|\tilde Y_{0,0}|^2$ along $XZ$ surface for different $\alpha=0,1,2,\ldots,10$; the dimensionless coordinates are $x = |\tilde
    Y_{0,0}(\theta,\phi)|^2 \sin\theta \cos\phi$ and $z = |\tilde Y_{0,0}(\theta,\phi)|^2 \cos\theta$ as a function of $\theta \in [0,\pi]$ and $\phi \in \{0,\pi\}$.  For
    $\alpha = 0$ the angular wave function is isotropic. For $\alpha>0$ the symmetry is broken, and as $\alpha$ increases the orbital is more elongated.
  }\label{fig:orbital00}
\end{figure}

\subsection{Low-lying states for large $\alpha$}

To solve the eigenproblem, as given by Eq.~\eqref{eigenprob}, for large $\alpha \gg 1$ it is better to return to Eq.~\eqref{angular}, and write $\tilde Y_{\ell,m}$ in the following form:
\begin{equation}
	\tilde Y_{\ell,m}(\theta,\phi) =  \Theta_{\ell,m}(\theta)\frac{e^{- i m \phi}}{\sqrt{2\pi}}.
\end{equation}
The problem of finding eigenvalues and eigenvectors of $\hat U$ reduces then to finding a solution for $\Theta_{\ell,m}(\theta)$ of the following equation:
\begin{equation}
\label{ThetaEQ}
	\bigg[\frac{ m^2}{\sin^2\theta} - \frac{1}{\sin\theta} \frac{\partial}{\partial \theta} \bigg(\!\! \sin\theta \frac{\partial}{\partial \theta} \bigg) - \alpha
          \cos\theta \bigg] \Theta_{\ell,m} \!=\!  \lambda_{\ell,m}\! \Theta_{\ell,m}.
\end{equation}
Note that $m$ enters only as $m^2$, so $\lambda_{\ell,m}$ and  $\Theta_{\ell,m}(\theta)$ depend only on~$|m|$.

First, we note, that for large $\alpha$, the solutions $\Theta_{\ell,m}(\theta)$ are localized around $\theta\approx 0$. This observation can be used to derive the
low-lying states, see Appendix~\ref{app-small-theta} for details, which are given by:
\begin{equation}
\label{evals}
	\lambda_{\ell,m} = - \alpha + \sqrt{2\alpha}(2\ell - |m| + 1). 
\end{equation}
The spectrum of $\hat U$ starts at $-\alpha + \sqrt{2\alpha}(|m|+1)$ and increases with $\ell$. The lowest lying eigenvalues are evenly distributed with interval $2\sqrt{2\alpha}$. The
eigenvalues increase with growing~$|m|$. We remark that the necessary condition for the validity of the presented approximation is $(2/\alpha)^{1/4}\ll1$ and $2\ell \ll
\sqrt{\alpha/2}+|m|$. This conditions are derived by analyzing the localization of the wave functions around $\theta\approx 0$.

The low-lying part of the spectrum, as indicated by Eq.~\eqref{evals}, is linear in $\ell$. In Fig.~\ref{fig:spectr} we present the spectrum of $\hat U$ for $\alpha =
3.65\times10^4$ for different values of $m$ calculated numerically by solving Eq.~\eqref{eigenprob}. This value of $\alpha$ roughly corresponds to collisions of KRb polar
molecule with $^{86}$Sr$^+$ or $^{87}$Rb$^+$ ions.  We observe at small values of $\ell$, that the spectrum exhibits linear behaviour, which is well described by
Eq.~\eqref{evals} for $m=0$ and $m=50$. For $m=100$ the slope of the linear dependence at small $\ell$ is slightly different than predicted by Eq.~\eqref{evals}. At large
values of $m$ the spectrum is no longer linear with $\ell$, and interestingly it becomes universal, not depending on the value of $m$.

%
%
\begin{figure}[htb!]
\includegraphics[clip, scale=0.55]{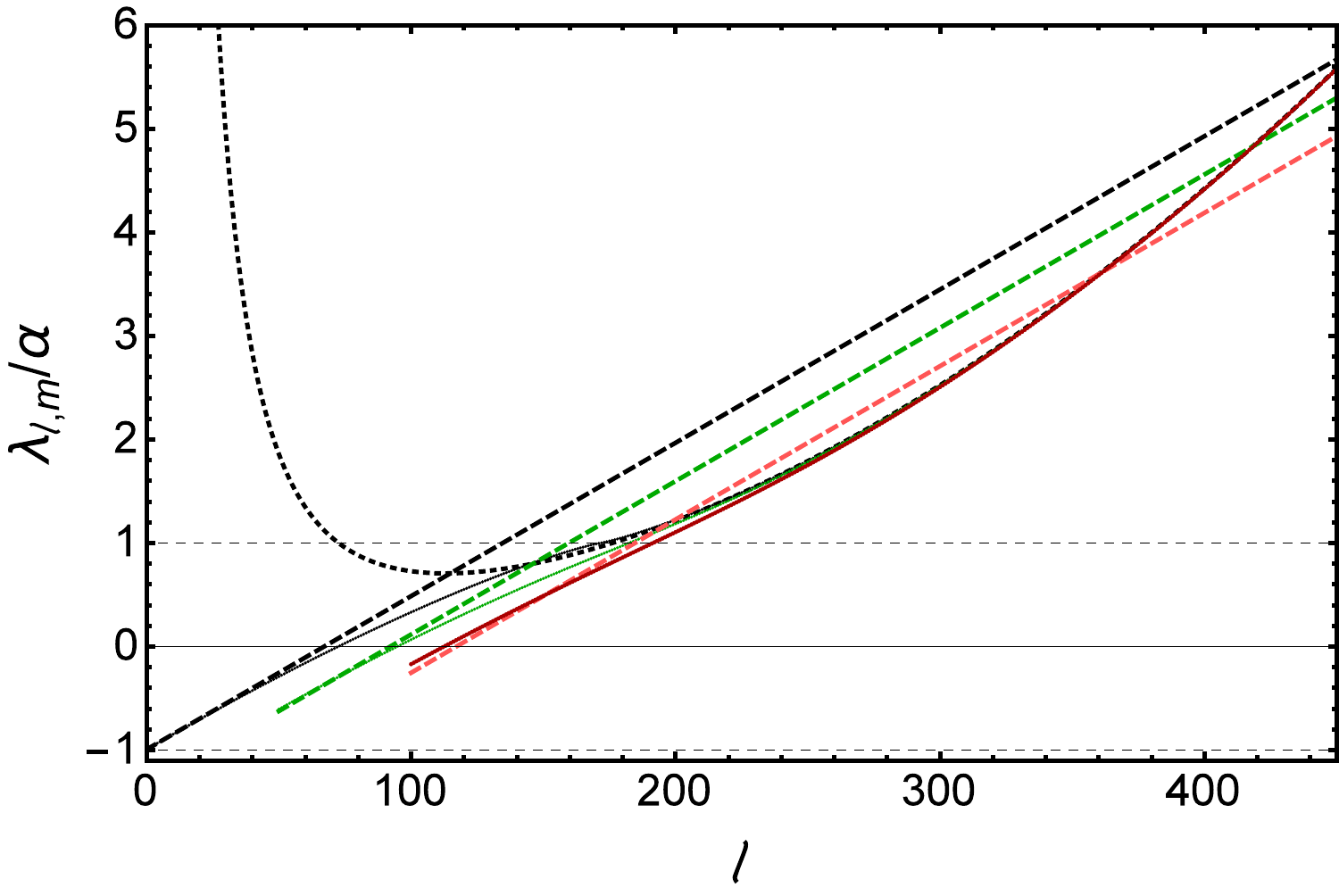}
  \caption{ 
    The spectrum $\lambda_{\ell,\,m}$ of $\hat U$ in units of $\alpha$ calculated numerically (solid lines) for $\alpha = 3.65\times 10^4$: $m=0$ (black), $m=50$
    (green) and $m=100$ (red). The dashed lines are the linear part of spectrum given by Eq.~\eqref{evals}. The dotted ``U'' shape (independent of $m$) is the
    asymptotic behavior given by the quasi-classical approximation Eq.~\eqref{asymptot}.  
  }\label{fig:spectr}
\end{figure}

Finally, we remark that the low-lying spectrum from Eq.~\eqref{evals}, can be rederived in a complementary way, by assuming that the coefficient
$c_{\ell,\ell'}^{(m)}$ are slowly varying functions of $\ell'$. For details, see Appendices~\ref{app-cont-l-intro} and~\ref{app-cont-l}.

%
%

\subsection{Quasi-classical approximation for large $\alpha$.}
\label{subsec:QuasiClassical}

To understand the behaviour of the higher-lying states of the angular part of the Schr\"odinger equation for $\alpha\gg1$, we employ the quasi-classical approximation.  Our
starting point is Eq.~\eqref{ThetaEQ}, which we rewrite with the function $\chi(\theta)$ defined by:
\begin{equation}
\label{chi}
	\chi_{\ell,\,m}(\theta) \equiv \Theta_{\ell,\,m}(\theta)\sqrt{\sin\theta} , 
\end{equation}
with the normalization $\int_{0}^\pi\!|\chi_{\ell,\,m}(\theta)|^2 d\theta = 1$.
This new function $\chi$ satisfies the following exact equation
\begin{equation}\label{chi_eq}
  - \chi_{\ell,\,m}''(\theta)\! +\! \tilde v(\theta)\chi_{\ell,\,m} = \varepsilon_{\ell,\,m} \chi_{\ell,\,m}(\theta),
\end{equation}
with the effective potential:
\begin{equation}
\label{tv}
	\tilde v(\theta) = \alpha(1- \cos\theta) + \frac{m^2-\frac14}{\sin^2\theta},
\end{equation}
where the eigenvalue $\varepsilon_{\ell,\,m}= \lambda_{\ell,\,m}+\alpha +\frac14$.
In this form, Eq.~\eqref{chi_eq} is the stationary Schr\"odinger equation for a particle moving in $m$-dependent potential $\tilde v(\theta)$ with energy~$\varepsilon_{\ell,m}$.

In Fig.~\ref{fig:wf} we plot the angular parts of the wave function $|\chi_{\ell,\, m}|^2$ as a function of $\theta$ for $m=20$ and $m=100$, and different $\ell=35, 115,
120, 200$.  These functions were calculated numerically from Eq.~\eqref{Ycl}, with the expansion coefficients $c_{\ell,\ell'}^{(m)}$ obtained by numerically solving the
recurrence relation~\eqref{eigenprob}. The solutions are rescaled and shifted in order to fit into the quasi-classical potential.

As we show in Appendix~\ref{subsec:QuasiClassical-app}, the behaviour of $\chi_{\ell,m}$ can be understood with help of the quasi-classical approximation, in which the wave function is
expressed as a superposition of the functions:
\begin{equation}
  \chi \sim k_\mathrm{cl}(\theta)^{-1/2} e^{\pm i \int^\theta k_\mathrm{cl}(\theta') d\theta'},
\end{equation}
where $k_\mathrm{cl}(\theta) = \sqrt{\varepsilon - v(\theta)}$ is the quasi-classical wave vector and $v(\theta)$ is given by $\tilde v(\theta)$ after applying 
the Langer correction (see Appendix~\ref{subsec:QuasiClassical-app}).

%
%
\begin{figure}[htb!]
	\includegraphics[clip, scale=0.55]{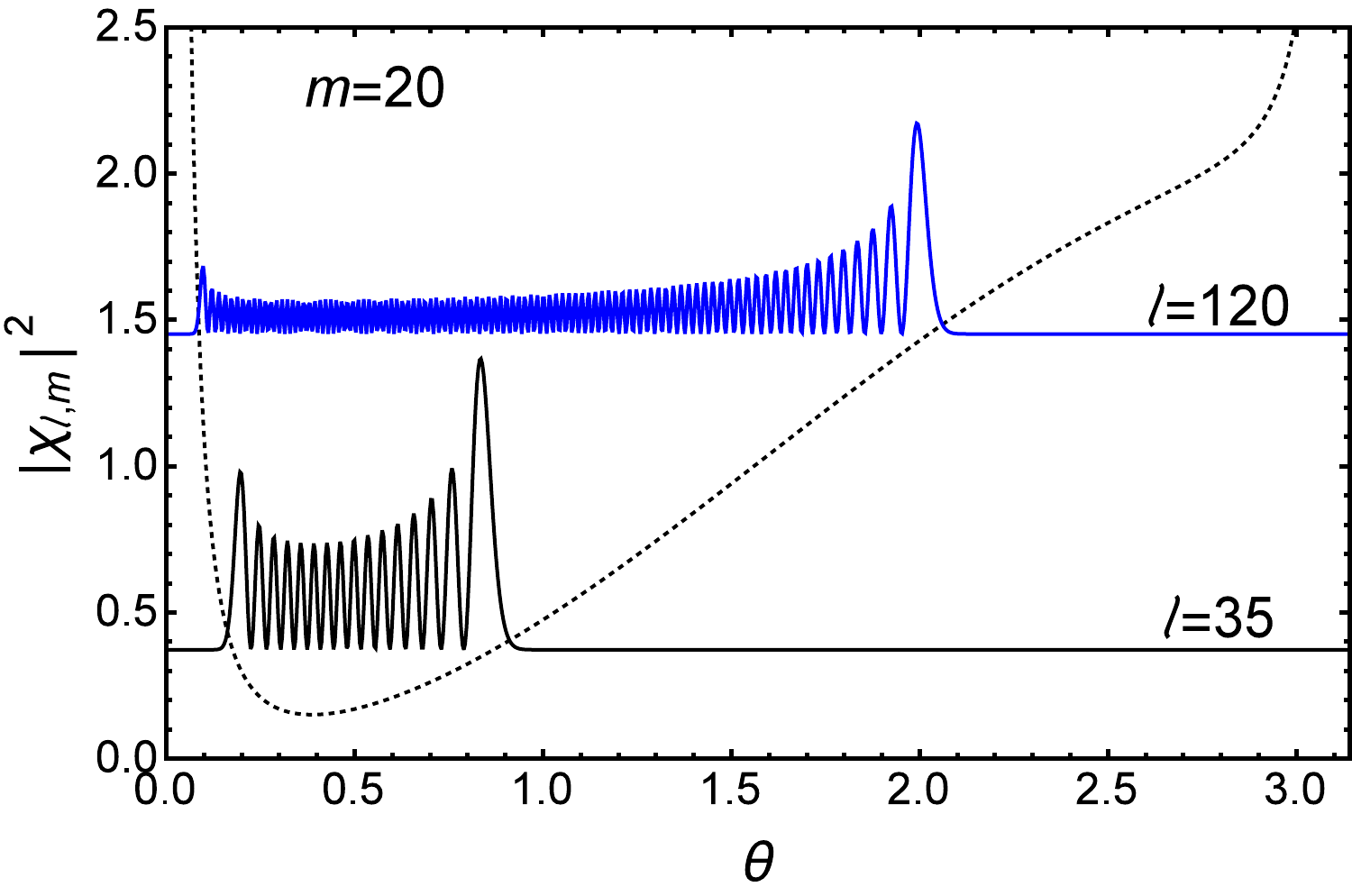}\\
	\includegraphics[clip, scale=0.55]{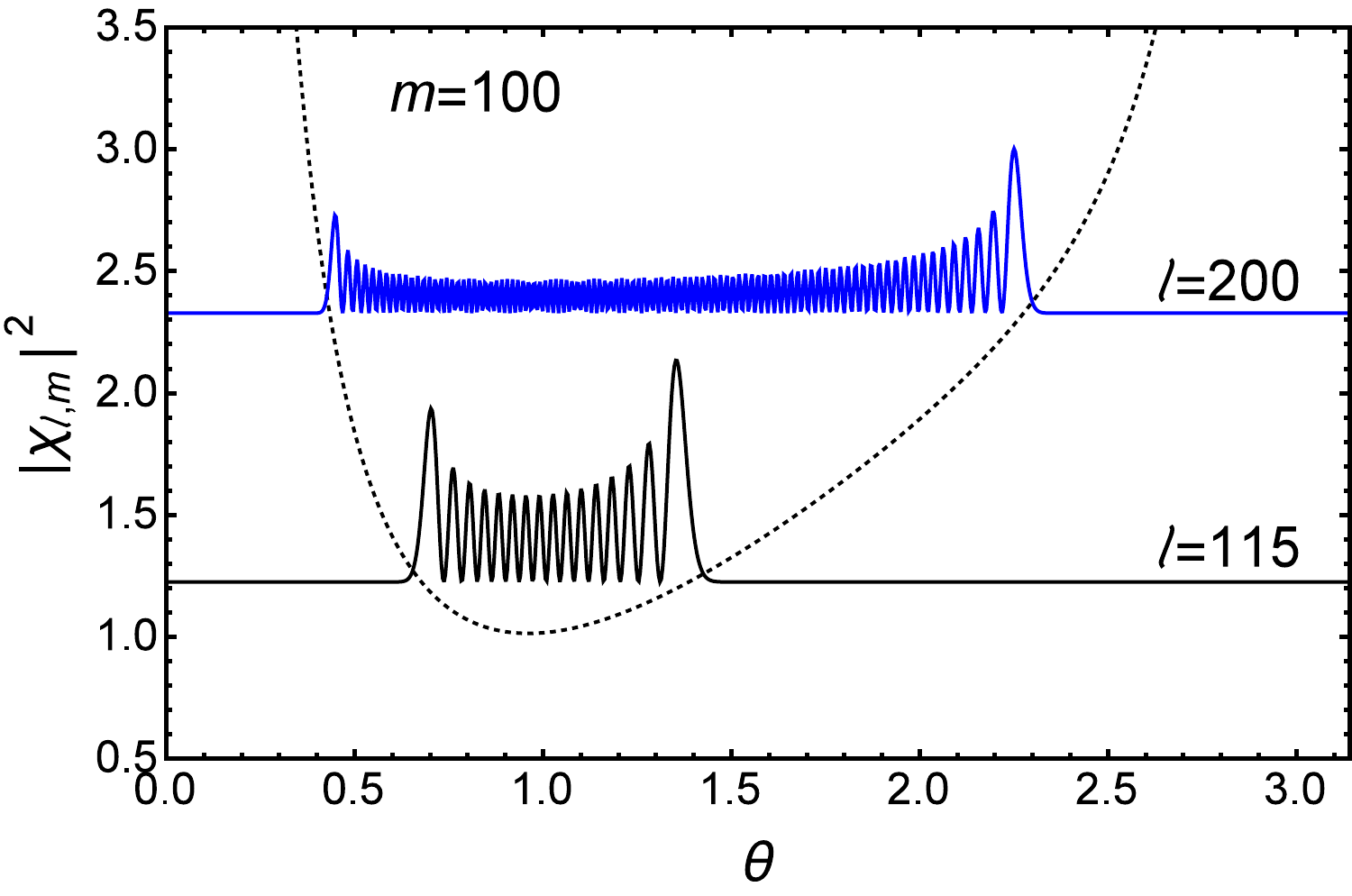}
	\caption{ 
		The angular part of the wave function $|\chi_{\ell,\,m}|^2$ for different values of $m$ and $\ell$.  Each wave function $|\chi_{\ell,\,m}|^2$ is shifted by its
		eigenvalue $\varepsilon_{\ell,\,m}$. The dotted line is the effective potential $v(\theta)$ calculated at given $m$.  The upper (lower) panel is calculated for $m=20$
		(m=$100$) and shows two wave functions indexed by $\ell=35$ and $\ell=120$ ($\ell=115$ and $\ell=200$). On both panels the lower and upper wave function has $n=15$ and
		$n=100$ nodes, respectively.}\label{fig:wf}
\end{figure}

The detailed analysis of these wave functions, as presented in Appendices~\ref{subsec:QuasiClassical-app} and~\ref{app-qcl}, shows that for small~$\ell$ we can recover the
spectrum given by Eq.~\eqref{evals}.  We also obtain a closed formula for large $\ell \gg \sqrt{\alpha}$, that is independent of $m$:
\begin{equation}\label{asymptot}
  \varepsilon_{\ell,m}  \xrightarrow[\ell \to \infty]{}  \alpha + \bigg(\ell+\frac12\bigg)^2 + \frac{\alpha^2}{8(\ell+1/2)^2}.
\end{equation}
This formula is depicted in Fig.~\ref{fig:spectr} with a dotted line. We observe that for large $\ell$ the eigenvalues corresponding to different $m$ indeed reduce to a single $m$-independent
curve given with a good approximation by Eq.~\eqref{asymptot}. 

Finally, we mention that the wave functions, as shown in Fig.~\ref{fig:wf}, can be very accurately described in the quasi-classical approximation. For a detailed analysis,
we refer the reader to Appendix~\ref{subsec:QuasiClassical-app}.

}


\section{Solution of the radial part}
\label{Sec:SolRad}
We turn now to the analysis of the radial part of the Schr\"odinger equation
\begin{equation}
\label{radial1}
\Big( - \frac{1}{r^2}\frac{\partial^2}{\partial r^2} - \frac2r \frac{\partial}{\partial r}  + \frac{\lambda_{\ell,m}}{r^2} \Big) R_{\ell,m}(r) = k^2 R_{\ell,m}(r).
\end{equation}
For the standard scattering problems, where the interaction potentials decays faster than $r^{-2}$, the centrifugal potential for all the partial waves except $s$-wave is
always repulsive. \revM{However, this is not the case for the ion-dipole potential, where some of the eigenvalues $\lambda_{\ell,m}$ can be negative}. Such negative values lead to
the attractive potential, which completely changes the properties of the wave functions at short distances.

The radial equation~\eqref{radial1} can be solved with Bessel functions of the first kind:
\begin{equation}
	R_{k,\ell,\,m}^{\pm}(r) = \sqrt{\frac{\pi}{2 k r}} J_{\pm \kappa}(kr),
\end{equation}
where $\kappa = \sqrt{\lambda_{\ell,\,m}+1/4}$ depends on the indices $\ell$ and $m$. For large distances $kr\gg 1$ we have
\begin{equation}
	R_{k,\ell,\,m}^{\pm}(r) \xrightarrow[k r \to \infty]{} \frac{1}{kr} \cos\bigg( kr \mp \frac{\kappa \pi}{2} - \frac{\pi}{4} \bigg),
\end{equation}
whereas for small $kr\ll 1$ we have
\begin{equation}
	R_{k,\ell,\,m}^{\pm}(r) \xrightarrow[k r \to 0]{} \sqrt{\frac{\pi}{2 kr}} \frac{1}{\Gamma(1\pm \kappa)} \bigg(\frac{kr}{2}\bigg)^{\pm \kappa}.
\end{equation}

The spectrum of the angular part of the wave function splits into two branches, which have different consequences in the radial part. We shift the eigenvalues as in the
previous section, introducing $\varepsilon_{\ell,\,m} = \lambda_{\ell,\,m} + \alpha + \frac{1}{4}$. For large $\alpha$, the spectrum of $\lambda_{\ell,m}$ starts at
$-\alpha$ (cf. Eq.~\eqref{evals}), hence, $\varepsilon_{\ell,m}\geq\frac14$.  The important parameter $\kappa$ that determines index of the Bessel function is
\begin{equation}
	\kappa_{\ell,\,m} = \sqrt{\varepsilon_{\ell,\,m} - \alpha}.
\end{equation}
For $\varepsilon_{\ell,\,m} \geqslant \alpha$, $\kappa_{\ell,\,m}$ is real and positive, so we may write $\kappa_{\ell,\,m} = |\kappa_{\ell,\,m}|$. Therefore, for large
$kr \gg 1$, the two solutions $R^{\pm}(r)$ are decaying as $\cos(kr - \phi^\pm_{\ell,m})/kr$, with the phase $\phi^{\pm}_{\ell,m} = (1\pm 2\kappa_{\ell,m}) \pi/4$. In
this regime the radial solutions decay at large distances as in the standard scattering problems, where the angular part is given by the usual spherical harmonics. For small $kr \ll 1$,
we have two solutions that have asymptotic behaviour $R^\pm(r) \propto (kr)^{\pm|\kappa| - 1/2}$.

It is instructive to analyse properties of the radial solutions in the limit of large angular momenta, which is equivalent to $\ell \to \infty$. Making use of the asymptotic formula \eqref{asymptot}, we obtain
\begin{equation}
\kappa_{\ell,\,m} \xrightarrow[\ell \to \infty]{}  \ell+\frac12 + \frac{\alpha^2}{16(\ell+1/2)^3}.
\label{kappalarge}
\end{equation}
The second term can be neglected for $\ell \gtrsim \alpha$, and in this case the radial solutions are identical as for standard scattering problem with short-range potentials
\begin{equation}
R_{k,\ell,\,m}^{\pm}(r) \xrightarrow[\ell \to \infty]{}  \sqrt{\frac{\pi}{2 k r}} J_{\pm \ell \pm \frac12}(kr),
\end{equation}

Different situation occurs for $\varepsilon_{\ell,\,m} < \alpha$. In that case $\kappa_{\ell,m}$ is purely imaginary, so $\kappa_{\ell,\,m} = i |\kappa_{\ell,\,m}|$. For small $kr\ll1$ we have an oscillating solutions,
$R^{\pm}(r) \propto (kr)^{-1/2}\exp(\pm i |\kappa_{\ell,m}|\log(kr))$. For large $kr\gg1$ we have the following asymptotic behaviour:
\begin{subequations}
\begin{eqnarray}
	R^{\pm}_{k,\ell,\,m}(r) \!\! \xrightarrow[k r \to \infty]{} & &
		\frac{
			e^{i(kr \!-\! \frac{\pi}{4})}e^{\pm \frac{|\kappa_{\ell,\,m}|\pi}{2}} 	}{2kr} + \nonumber \\
			& & + \frac{ e^{-i(kr \!-\! \frac{\pi}{4})}e^{\mp \frac{|\kappa_{\ell,\,m}|\pi}{2}}
		}{2kr}.
\end{eqnarray}
\end{subequations}

\section{Reactive collisions in the universal regime}
\label{Sec:Univ}
In this section we employ the solutions of the radial equation to analyze reactive collisions in the universal regime \cite{Idziaszek2010,Idziaszek2010a}. To this end, we
adopt the approach within QDT, which is based on the parametrization of the wave function at small distances, where, with a good approximation, it does not depend either
on the energy or on the angular momentum of the collision. Specifically, for the solution with imaginary $\kappa$, i.e., when $\lambda_{\ell,m}+1/4 < 0$, the short
distance wave function oscillates for $r \to 0$, and
\begin{equation}
  R(r) \approx \frac{c_1}{\sqrt{r}} e^{+ i |\kappa| \textrm{ln}\frac{kr}{2}} + \frac{c_2}{\sqrt{r}} e^{- i |\kappa| \textrm{ln}\frac{kr}{2}}.
\end{equation}
The parts of the wave function with + and - describe the outgoing and incoming probability currents, respectively. Within QDT we parameterize
\begin{equation}
  c_1 = \frac{1-y}{1+y} e^{i \phi}, \quad c_2 = e^{-i \phi}.
\end{equation}
In general both the short-range phase $\phi$ and the parameter $y$, which describes the probability of the short-distance reaction, can depend on $\ell$ and $m$. Below, 
we analyze the simplest possible case, i.e., we assume that $y=1$. In this universal limit, the scattering properties do not depend on the short-distance phase $\phi$. 
The scattering matrix is given by
\begin{equation}
\label{s_uni}
  S_{\ell,m} = i (-1)^\ell e^{- \pi |\kappa_{\ell,m}|},
\end{equation}
and is valid for $\lambda_{\ell,m}+1/4 <0$. 

In the other regime, if $\lambda_{\ell,m}+1/4 >0$, $\kappa$ is real and positive, and then the wave function
at short distances takes the form
\begin{equation}
  R(r) \approx d_1 \frac{(k r)^{+\kappa} }{\sqrt{r}} + d_2 \frac{(k r)^{- \kappa}}{\sqrt{r}}.
\end{equation}
Physically meaningful solution is obtained when assuming $d_2 = 0$, which, after a straightforward calculation, leads to the following $S$ matrix
\begin{equation}
  S_{\ell,m} = i (-1)^\ell e^{-i \pi \kappa_{\ell,m}}.
  \label{Slm2}
\end{equation}
Taking into account \eqref{kappalarge}, in the limit of large $\ell$ we obtain 
\begin{equation}
S_{\ell,\,m} \xrightarrow[\ell \to \infty]{}  \exp \left( -i \frac{\pi \alpha^2}{16(\ell+1/2)^3} \right).
\end{equation}

Now, we can proceed to the evaluation of the reactive collision rate given by Eq.~\eqref{Kre_eq}. The partial waves for $\lambda_{\ell,m}>-1/4$ do not contribute to
${\cal K}^\mathrm{re}$, since $|S_{\ell,m}| = 1$ (cf. Eq. \eqref{Slm2}). Physically, this corresponds to the scattering on the repulsive potential, where the particles do
not approach the core region $r =0$, where the reaction takes place. Correspondingly, the only contribution to ${\cal K}^\mathrm{re}$ comes from the partial waves with
$\lambda_{\ell,m}<-1/4$.

If $\alpha\gg1$ we can get an estimate of the reactive collision rate, since most of the contributing $\kappa$ are large, and according to Eq.~\eqref{s_uni}, $S_{\ell,m} \approx 0$. The reactive rate is then given by
\begin{equation}
\label{Ksum}
  {\cal K}^\mathrm{re} \approx \frac{\pi \hbar}{\mu k_i} \sum_{|m| \leqslant m^*} \sum_{|m| \leqslant \ell \leqslant \ell_m^*} 1,
\end{equation}
where $\ell_m^*$ is the maximum $\ell$ for which $\lambda_{\ell,m} < 0$, and $m^*>0$ is the largest $m$ for which negative $\lambda_{\ell,m}$ do exist. 
In the first approximation, as can be inferred from Eq.~\eqref{evals},  $m^* = \sqrt{\alpha/2}$ and $\ell_m^* = (m^* + |m|)/2$. This leads to
\begin{equation}
\label{Kre_classical}
  {\cal K}^\mathrm{re} \approx \frac{\pi \hbar}{\mu k_i} \frac{\alpha}{4}.
\end{equation}
Since the number of states, for which the particles can collide and react, is proportional to $\alpha$, the reactive rate is also proportional to $\alpha$. We note that the reactive rate depends on the energy as $1/k \sim1/\sqrt{E}$. This behaviour agrees with the prediction for the power-law potentials $V(r) = - C_n/r^n$ \cite{Jachymski2013}, i.e., 
\begin{equation}
{\cal K}^\mathrm{re} \stackrel{E\to\infty}{\longrightarrow}   g \frac{h}{2 \mu k} P^\mathrm{re} \frac{n}{2} \left( \frac{E/E_n}{\frac{n}{2}-1}\right)^{(n-2)/n},
\label{KLang}
\end{equation}
where in the limit of $n \to 2$ we obtain
\begin{equation}
{\cal K}^\mathrm{re} \stackrel{E\to\infty}{\longrightarrow}   \frac{\pi \hbar}{\mu k} \frac{2 \mu C_2}{\hbar^2},
\label{KLang1}
\end{equation}
substituting for probability of reaction $P^\mathrm{re} =1$ and $g=1$ for distinguishable particles.

Alternatively, the formula from Eq.~\eqref{Kre_classical}, can be derived based on classical considerations. To be specific, one should solve the classical equations of motion and count only the contribution to the reactive rate from the trajectories that fall to the scattering center. By calculating contribution from such trajectories, one can reproduce exactly the formula~\eqref{Kre_classical}.

\begin{figure}[htb!]
\includegraphics[clip, width=\columnwidth]{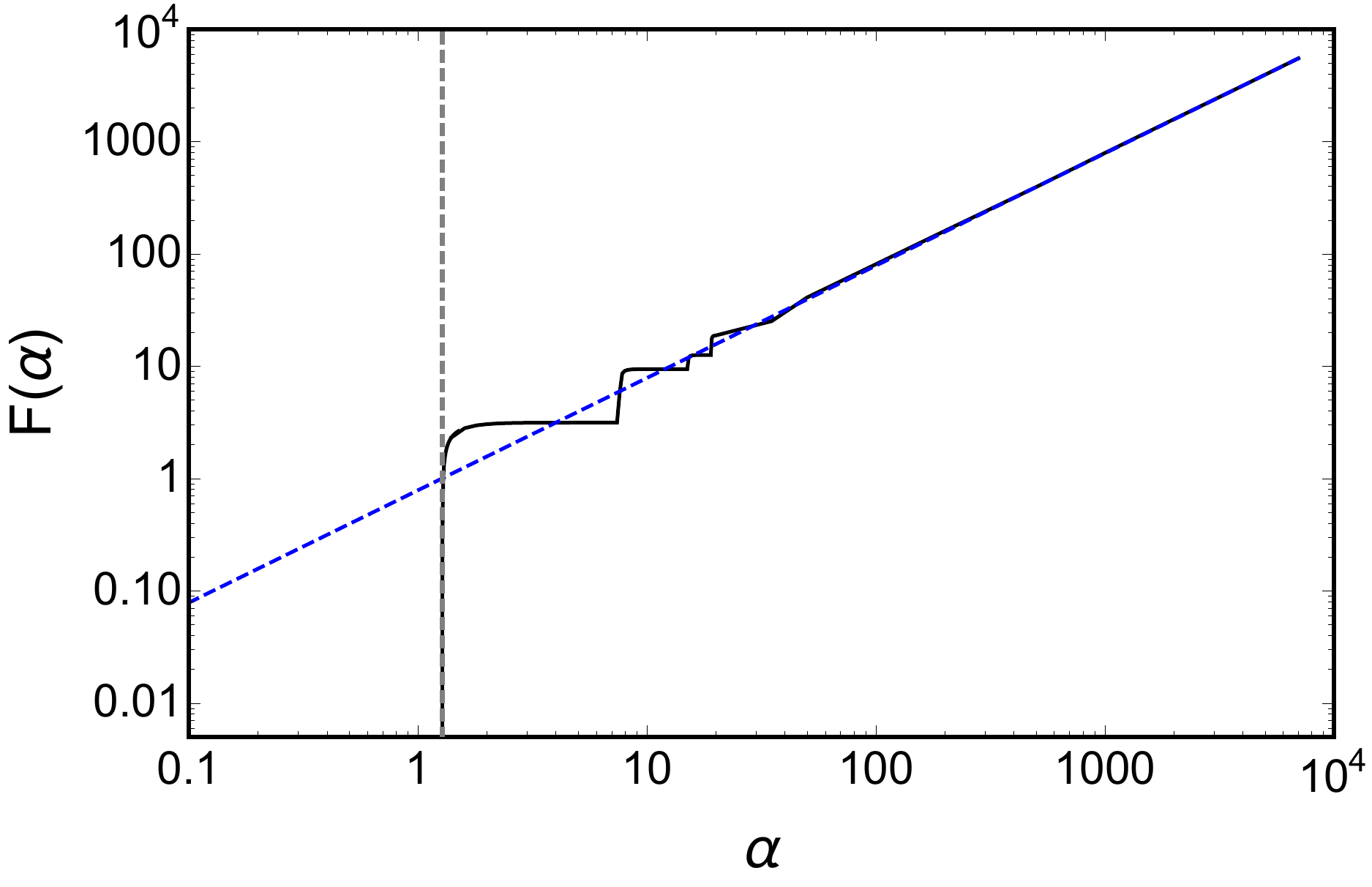}
  \caption{ The function $F$ that determines the universal collisional rate $K^\mathrm{re}$ in the universal reactive case, see Eq.~\eqref{Kre_F}. 
    The solid black curve with steps is the quantum mechanical rate determined
    numerically with Eq.~\eqref{Kre_eq}, where the $S$-matrix is taken from Eq.~\eqref{s_uni}. 
    The dashed blue straight line is the classical case given by Eq.~\eqref{Kre_classical}.
    Dashed vertical line displays $\alpha_\mathrm{cr} = 1.279$ for which the first state enters into the regime where $\lambda_{\ell,m} < -1/4$.
  }\label{fig:F}
\end{figure}

In general, the universal collisional rate will have the following form
\begin{equation}
\label{Kre_F}
  {\cal K}^\mathrm{re} = \frac{\hbar}{\mu k_i} F(\alpha),
\end{equation}
where $F$ is a universal function depending only on the dipole-ion interaction strength $\alpha$. In Fig.~\ref{fig:F} we plot the function $F(\alpha)$ that we calculate
numerically. The function exhibits steps, when new states enter below the threshold $\lambda_{\ell,m} < -1/4$. Since a finite value of $\alpha$ is needed to generate such
a state, the function $F$ is nonzero for $\alpha \gtrsim \alpha_\mathrm{cr} = 1.279$. The threshold $\alpha_\mathrm{cr}$ is the maximal value of $\alpha$ for which the radial
Schr\"odinger equation has \revM{at least one solution that is} finite at $r \to 0$ \cite{Mittleman1965}. For large values of $\alpha$ the function $F$ approaches the classical limit given
by $\pi \alpha/4$. In particular, for $\alpha \gtrsim 200$, the error is smaller than 2\%.

%
%

\section{Conclusions}
\label{Sec:Concl}

In this work, we have investigated a reactive scattering in the ion-dipole potential. First, we have introduced the modified spherical harmonics, describing angular wave
functions in the presence of an anisotropic ion-dipole interaction, and their corresponding eigenvalues. For large values of $\alpha$, \revM{which quantifies the strength
  of the potential, } we have derived a number of analytical approximations based on an expansion at small angles, continuous-$\ell$ quantum number approximation and the
semi-classical method. 

We have shown that the introduction of the modified spherical harmonics, requires some modifications of the formulas for the elastic and reactive total cross sections and
collision rates. We have investigated the properties of the radial solutions, which are given in terms of Bessel functions, analysing two different regimes, corresponding
basically to attractive and repulsive long-range interaction potentials.

Finally, we have calculated the collision rates for the reactive scattering in the universal regime, where the short-range reaction probability is equal to unity. In such
a case, it is not necessary to introduce any other phase parameters describing the short-range behaviour of the wave function.  
\revM{
We have shown that this rate scales inversely proportional to the square-root of the energy times a function $F(\alpha)$ that depends only on the strength of the
potential. This universal function $F(\alpha)$ is nonzero only beyond the threshold $\alpha_\mathrm{cr} \approx 1.279$. It exhibits quantum step-like dependence for
$\alpha \sim 1$ and it approaches the classical limit for large~$\alpha$. It would be interesting to see that the inlcusion of a short-range potential might lead to a 
quantum reflection at intermediate distances and to shape resonances manifested as additional structures~in~$\mathcal{K}^\mathrm{re}$.}

\revM{ 
In the ultracold regime the knowledge of the long-range part of the potential is not sufficient to determine the phase shift of the wave function at large distances and
the scattering properties crucially depend on the short-range interaction. The ion-dipole potential does not possess any characteristic energy nor length scale. Moreover,
in the regime of large parameter $\alpha$, there is no quantum reflection process, and the collision dynamics is semiclassical to a large degree. In that sense the
ion-dipole potential is transparent for inelastic or reactive scattering, modifying only the long-range angular properties of the wave function. 
Therefore, it is crucial to include the short-range part in potential in the considerations.
	
In the case very relevant to the current experiments, i.e collisions of ground state polar molecules build-up of two-alkali metal atoms with alkaline earth metal ions,
the next order dispersion term in the expansion of the long-range potential, originates from the off-resonant ion-induced-dipole interaction, neglecting ion-quadrupole
interaction, which for the considered systems is rather weak. At large distances, in the second-order perturbation theory, the next order term behaves as $-C_4/r^4$ with
$C_4=d^2/(6 B)$ \cite{Idziaszek2011a}, where $d$ is a permanent dipole moment of the molecule, and $B$ is its rotational constant. With this long-range polarization
potential one can associate characteristic range $R^\ast=(2 \mu C_4)^{1/2}/\hbar$ and the energy $E^\ast=\hbar^2/(2 \mu (R^\ast)^2)$] \cite{Idziaszek2009a}. For the
considered systems the characteristic energy ranges from about 50 nK for LiNa--$^9$Be$^+$ to 5 pK for LiCs--$^{174}$Yb$^+$ system, setting the height of the $p$-wave
centrifugal barrier. In that sense going to the quantum regime of the scattering dominated only by the lowest partial wave collisions could be extremely difficult in
comparison to neutral or hybrid atom-ion systems. At the same time, characteristic range of this potential is relatively large, ranging from 0.8 $\mu$m (LiNa--Be$^+$) up
to 27 $\mu$m (LiCs--Yb$^+$).
}

\section{Acknowledgements}
This work was supported by the Polish National Science Center project 2014/14/M/ST2/00015.

\appendix

%
%

\revM{
\section{Approximation schemes for the  angular part for $\alpha \gg 1$}
\label{App:SolAng}
}

\revM{
In this appendix we present details on the approximate methods of evaluating the angular part of the Schr\"odinger equation in the limit $\alpha\gg1$.

Our starting point is the equation for $c^{(m)}_{\ell,\ell'}$ (or equivalently on  $\tilde Y_{\ell,m}$) in the form of Eq.~\eqref{eigenprob}. Denoting now the eigenvalues of $\hat l^2$ by $V_\ell = \ell(\ell+1)$, and the contribution from the dipole by $D_\ell = - \alpha \beta_{m,\ell}$, the matrix $\hat U$ in the basis of standard spherical harmonics  $Y_{\ell,m}$ takes the following form:
\begin{widetext}
\begin{equation}
\label{matU}
	\hat U =
	\begin{pmatrix}
		V_{m} 		& D_{m+1} 		& \cdot 				& \cdot  			& \cdot  & \cdot 		& \cdot 	& \cdot 		& \ \cdot\  & \ \cdot\ \\
		D_{m+1} 	&	V_{m+1} 		& D_{m+2} 			& \cdot  			& \cdot  & \cdot 		& \cdot 	& \cdot 		& \cdot & \cdot \\
		\cdot 		& D_{m+2} 		& V_{m+2}				& D_{m+3}  		& \cdot  & \cdot 		& \cdot 	& \cdot 		& \cdot & \cdot \\
		\cdot 		& \cdot				& \ddots 				& \ddots  		& \ddots & \cdot 		& \cdot 	& \cdot 		& \cdot & \cdot \\
		\cdot 		& \cdot				& \cdot 				& D_{\ell-1}  		& V_{\ell-1}& D_{\ell} 		& \cdot 	& \cdot 		& \cdot & \cdot \\
		\cdot 		& \cdot				& \cdot 				& \cdot 	 		& D_\ell		 & V_\ell	 		& D_{\ell+1}	& \cdot 		& \cdot & \cdot \\
		\cdot 		& \cdot				& \cdot 				& \cdot  			& \cdot  & D_{\ell+1}	& V_{\ell+1}	& D_{\ell+2} 	& \cdot & \cdot \\
		\cdot 		& \cdot				& \cdot 				& \cdot  			& \cdot  & \cdot 		& \ddots 	& \ddots 		& \ddots& \cdot
		\end{pmatrix}.
\end{equation}
\end{widetext} 

This form of the matrix is used in numerical simulations, where we additionally impose a cutoff $\ell_\mathrm{max}$.
}

%
%
\revM{


\subsection{Expansion for small $\theta$}
\label{app-small-theta}

The numerical calculations show that for large values of $\alpha$, the lowest orbitals $\tY{\ell}{m}$ are localized around $\theta\approx 0$. In such a case one can expand
Eq.~\eqref{angular} around that point, and obtain approximate form of angular orbitals analytically. 

To this end, we expand Eq.~\eqref{ThetaEQ} for small $\theta$ substituting: $\cos\theta \approx 1 -\frac12 \alpha^2$ and $\sin \theta
\approx \theta$. This leads to
\begin{equation}
	\bigg[
	- \frac{\partial^2}{\partial \theta^2}
	- \frac{1}{\theta} \frac{\partial}{\partial \theta}
	+\frac{ m^2}{\theta^2}
	- \alpha  + \frac12 \alpha \theta^2 \bigg] \Theta_{\ell,m}(\theta) =
	\lambda_{\ell,m}\Theta_{\ell,m}(\theta).
\end{equation}
The solutions that satisfy appropriate boundary conditions, i.e., is finite for $\theta=0$ and vanish for large $\theta$, are given in terms of confluent hypergeometric function:
\begin{equation}
\label{evecs}
	\Theta_{\ell,m}(\theta)\!\! =\!\! \mathcal{N}_{\ell,m}^{\frac12} \theta^{|m|} e^{- \frac{1}{2}\sqrt{\frac{\alpha}{2}} \theta^2 }\, {}_1F_1\bigg(-n, |m|+1,
        \sqrt{\frac{\alpha}{2}} \theta^2 \bigg),
\end{equation}
where the quantum number $n=\ell-|m|$=0, 1, 2, \ldots, indexes the number of nodes of the angular wave function.
The resulting eigenvalues are given by Eq.~\eqref{evals}. For completeness, we give here the normalization constant, which is given by
\begin{equation}
	\mathcal{N}_{\ell,m} = \frac{ 2 (\alpha/2)^{(|m|+1)/2} }{ (|m|)! } \binom{\ell}{|m|}.
\end{equation}

From the form of solution \eqref{evecs} it can be seen that the function $\Theta_{\ell,m}(\theta)$ is negligible for $\theta \gg (2/\alpha)^{1/4}$, and combining this
with the condition $\theta\ll 1$ we obtain the necessary condition for the validity of the presented approximation
\begin{equation}
(2/\alpha)^{1/4}\ll1.
\end{equation}
If this parameter is small, the approximation that led to \eqref{evecs} and \eqref{evals} is applicable. This condition, however, is not
sufficient, because for large $\ell$, the wave function extends to larger $\theta$, and we violate the condition $\theta \ll 1$.  From quasi-classical considerations (see subsection~\ref{subsec:QuasiClassical}) we obtain that the region of nonvanishing wave function is of the order of $\theta \lesssim \sqrt{(1+\lambda_{\ell,\,m}/\alpha)}$.  This gives the required
condition for $\ell$
\begin{equation}
2\ell \ll \sqrt{\alpha/2}+|m|.
\end{equation}

%
%
\subsection{Continuous-$l$  approximation.}
\label{app-cont-l-intro}

In this section we exploit the fact that for large $\alpha\gg1$ the coefficients $c_{\ell,\ell'}^{(m)}$ in Eq.~\eqref{Ycl}, which enter the recurrence relation~\eqref{eigenprob},
change smoothly with $\ell$. The details of the derivation can be found in Appendix~\ref{app-cont-l}, here, we merely state the final results.

First, we introduce the small parameter of the expansion, which is $\epsilon = \alpha^{-1/4}$.
Then we drop indices $\ell$ and $m$, rename $\lambda_{\ell,\,m}$ to $\lambda$ and write $c_{\ell,\,l}^{(m)}$ as $c(l)$. Next, we introduce new variable  $x = \epsilon l$ and define $\tilde c(x) = \tilde c(\epsilon l) \equiv c(l)$, and new parameter $\tilde \lambda$ by the relation: $\tilde \lambda = \lambda \epsilon^4$, where $\tilde \lambda$ is of the order of unity.
After expanding in small parameter $\epsilon$, Eq.~\eqref{eigenprob} takes the following form (with higher order terms being neglected):
\begin{equation}
\label{eq_pos}
	- \frac12 \tilde c''(x) + \bigg( 
	\frac{m^2-1/4}{2 x^2} +x^2 \bigg) \tilde c(x)  = \frac{\tilde\lambda+1}{\epsilon^2}\tilde c(x).
\end{equation}
The solution that is finite both at $x \to 0$ and for large $x$ is given in terms of the confluent hypergeometric function:
\begin{equation}
	\tilde c^{(n)}(x) \propto e^{-\frac{x^2}{\sqrt2}}x^{|m|+1/2} {}_1F_1(-n, |m|+1, \sqrt{2}x^2),
\end{equation}
where $n=0,1,2,3,\ldots$ is the quantum number labelling the solutions, and the eigenvalues are given by
\begin{equation}
\label{evapprox}
	\lambda_n = - \alpha  + \sqrt{2\alpha}(2n +|m|+ 1).
\end{equation}
We notice that eigenvalues in Eq.~\eqref{evapprox} are identical as in Eq.~\eqref{evals}.

%
%
\subsection{Quasi-classical approximation.}
\label{subsec:QuasiClassical-app}

In the quasi-classical \revM{treatment of Eq.\eqref{chi_eq}}
we introduce left and right classical turning points, denoted further by $\theta_L$ and $\theta_R$, respectively, defined by $k_\mathrm{cl}(\theta_L)=k_\mathrm{cl}(\theta_R)=0$, and $\theta_L < \theta_R$. The classical wave vector $k_\mathrm{cl}$ is given by
\begin{equation}
\label{kcl}
	k_\mathrm{cl}(\theta) = \sqrt{\varepsilon - v(\theta)}.
\end{equation}
Applying the quasi-classical method to the radial Schr\"odinger equation requires inclusion of the so-called Langer correction \cite{Langer1937}, which basically boils down to dropping of $1/4$  in the second term of  Eq.~\eqref{tv}
\begin{equation}
\label{v}
	v(\theta) = \alpha(1- \cos\theta) + \frac{m^2}{\sin^2\theta}.
\end{equation}

Within the quasi-classical approximation, the wave functions for the eigenstates in the classically accessible region are given by:
\begin{equation}
\label{CLacc}
\chi(\theta) = \frac{C}{\sqrt{k_\mathrm{cl}(\theta)}} \cos\bigg( \int_\theta^{\theta_R}\! k_\mathrm{cl}(\theta')d\theta' - \frac\pi4 \bigg),
\end{equation}
for $\theta_L < \theta < \theta_R$, and far from the ends of that interval. In the classically inaccessible region, the wave function are given by
\begin{equation}
\label{CLinaccR}
\chi(\theta) = \frac{C}{2\sqrt{|k_\mathrm{cl}(\theta)|}} \exp\bigg( - \int_{\theta_R}^{\theta}\! |k_\mathrm{cl}(\theta')| d\theta' \bigg),
\end{equation}
for $\theta > \theta_R$, and
\begin{equation}
\label{CLinaccL}
\chi(\theta) = \frac{(-1)^n C}{2\sqrt{|k_\mathrm{cl}(\theta)|}} \exp\bigg( - \int_{\theta}^{\theta_L}\! |k_\mathrm{cl}(\theta')| d\theta' \bigg),
\end{equation}
for $\theta < \theta_L$. 
For completeness we give the expression for the normalization constant,
$C = \big(\frac12 \int_{\theta_L}^{\theta_R} k_\mathrm{cl}^{-1}(\theta')d\theta'\big)^{-1/2}$.

The eigenvalues $\varepsilon$ can be obtained from the Bohr-Sommerfeld's quantization rule, given by:
\begin{equation}
\label{BS}
	\int_{\theta_L}^{\theta_R} \!\! k_\mathrm{cl}(\theta') d\theta' = \bigg(n+\frac12\bigg)\pi,
\end{equation}
where $n$ is a positive integer number indexing the $n$-th eigenvalue in the potential $v(\theta)$.  \revM{Note that $n = \ell - |m|$ and from the method we obtain
$\varepsilon_{\ell,m}$.  We have checked numerically that inclusion of the correction $-1/4$ in the definition of $\lambda_{\ell,m}$ significantly improves the accuracy
of the approximate solutions.}

The procedure of solving the eigenproblem for the angular part, defined by equation~\eqref{ThetaEQ}, is now straightforward.
We first find the shifted eigenvalues $\varepsilon_n$ indexed by a non-negative integer $n$ by solving Born-Sommerfeld's quantization conditions~\eqref{BS}.
The wave functions are then given by equations:~\eqref{CLacc} -- classically accessible region,
\eqref{CLinaccR} -- right classically inaccessible region, and \eqref{CLinaccL} -- left classically inaccessible region.

In Appendix~\ref{app-qcl} we derive formulas for the eigenvalues that are obtained from the Bohr-Sommerfeld's quantization rule~\eqref{BS} for small $|m|$.   For $0\leqslant \varepsilon < 2\alpha$ we obtain the following equation determining the eigenvalues $\varepsilon$
\begin{equation}
\label{BS_1}
\int_{0}^{\theta_R} \!\! \sqrt{\varepsilon - \alpha(1-\cos\theta')} d\theta' = \bigg(2n+|m|+1\bigg)\frac{\pi}{2},
\end{equation}
with $\theta_R = \arccos(1-\varepsilon/\alpha)$. In the other regime, for the eigenvalues $\varepsilon\geqslant2\alpha$ we arrive at the following formula:
\begin{equation}
\label{BS_2}
\int_{0}^{\pi}\!\!\! \!\! \sqrt{\varepsilon \!-\! \alpha(1\!-\!\cos\theta')} d\theta' \!=\! \bigg(\!n\!+\!|m|\!+\!\frac12\bigg)\pi \!=\! \bigg(\!\ell + \frac12\bigg)\pi.
\end{equation}

In Fig.~\ref{fig:sp1} we display the spectrum $\lambda_{\ell,\,m}$ for $m=20$ and $\alpha=3.65\times 10^4$. We compare the numerically calculated values from
Eq.~\eqref{eigenprob} to the ones obtained by solving the quasi-classical quantization rule, Eq.~\eqref{BS}. We also display eigenvalues obtained by approximate
quasi-classical quantization conditions, Eqs.~\eqref{BS_1} and~\eqref{BS_2}. All the solution are in very good agreement with the exact numerical result. The relative
error , which is shown in the inset, remains below $10^{-3}\%$ for the full quasi-classical formula and is larger for the quasi-classical solution without inclusion of
the Langer correction, and for asymptotic formula~\eqref{asymptot}.

%
%
\begin{figure}[htb!]
	\includegraphics[clip, scale=0.40]{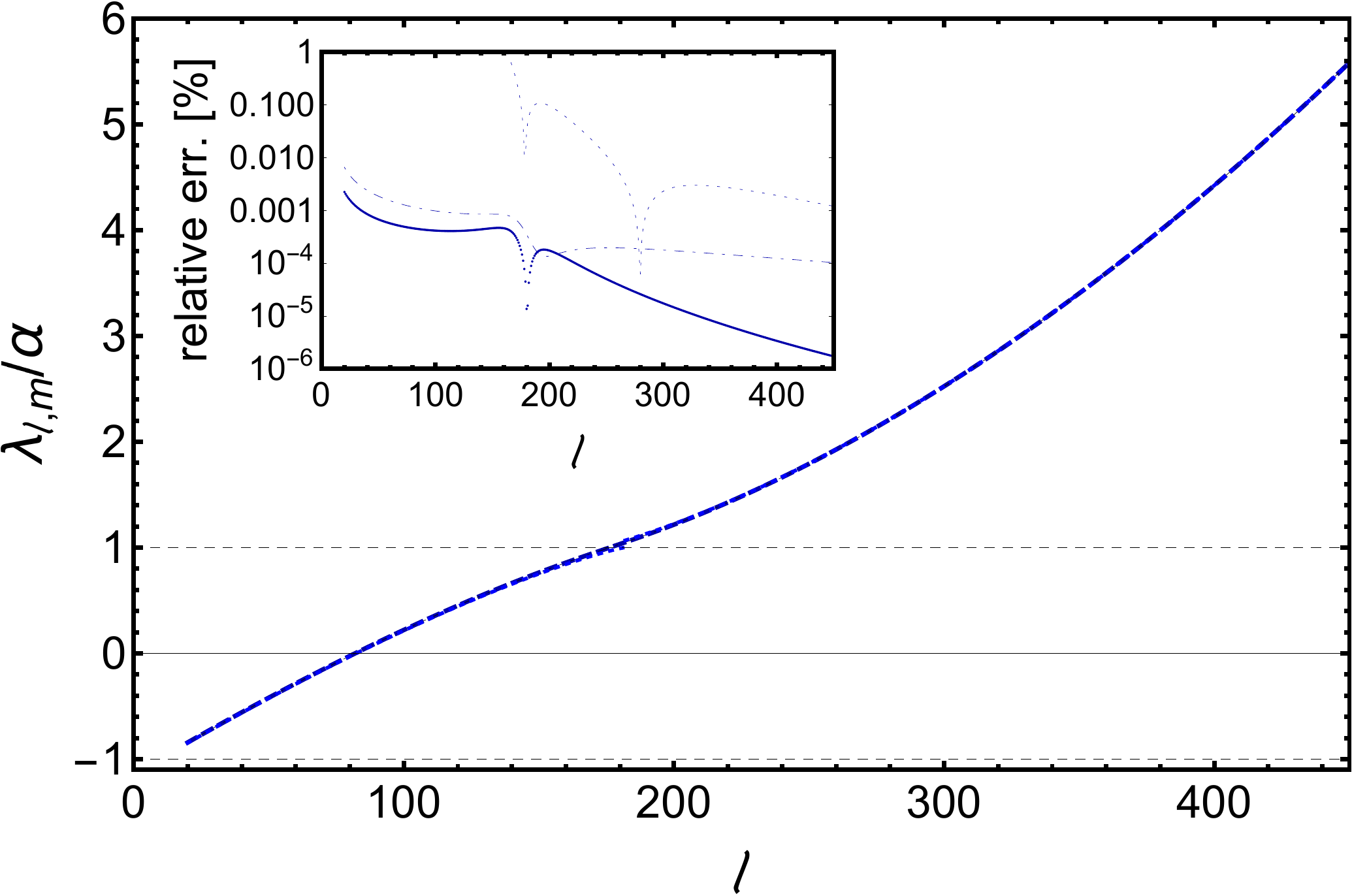}
	\caption{
		The spectrum $\lambda_{\ell,\,m}$ of $\hat U$ for $m=20$ in units of $\alpha$ calculated numerically for $\alpha = 3.65\times 10^4$ (dashed blue line). Dot-dashed
		black line (coincides with full numerical dashed blue) is the spectrum obtained within quasi-classical approximation (see Eq.~\eqref{BS}). Dotted blue lines (coinciding with the previous two) represent solutions given by Eqs.~\eqref{BS_1} and~\eqref{BS_2}. The inset shows the
		absolute relative error between solutions given by numerical diagonalization and: blue dots -- quasi-classical approximation given by Eq.~\eqref{BS} (smallest relative
		error), dot-dashed -- quasi-classical approximation with omitted shift 1/4, i.e., $\lambda = \varepsilon - \alpha$ (medium relative error), and dotted line - the asymptotic form
		given by Eq.~\eqref{asymptot} (largest relative error).  
	}\label{fig:sp1}
\end{figure}

In Fig.~\ref{fig:sp2} we show the spectrum and relative errors in the case $m=100$. Here the condition $|m| \ll \alpha$ is not satisfied, and the potential $v(\theta)$ is strongly affected 
by the presence of the term $m^2/\sin^2\theta$ in Eq.~\eqref{v}. As a consequence it cannot be neglected, and so the formulas given by Eqs.~\eqref{BS_1} and \eqref{BS_2} are not as accurate as for $m=20$.

%
%
\begin{figure}[htb!]
	\includegraphics[clip, scale=0.40]{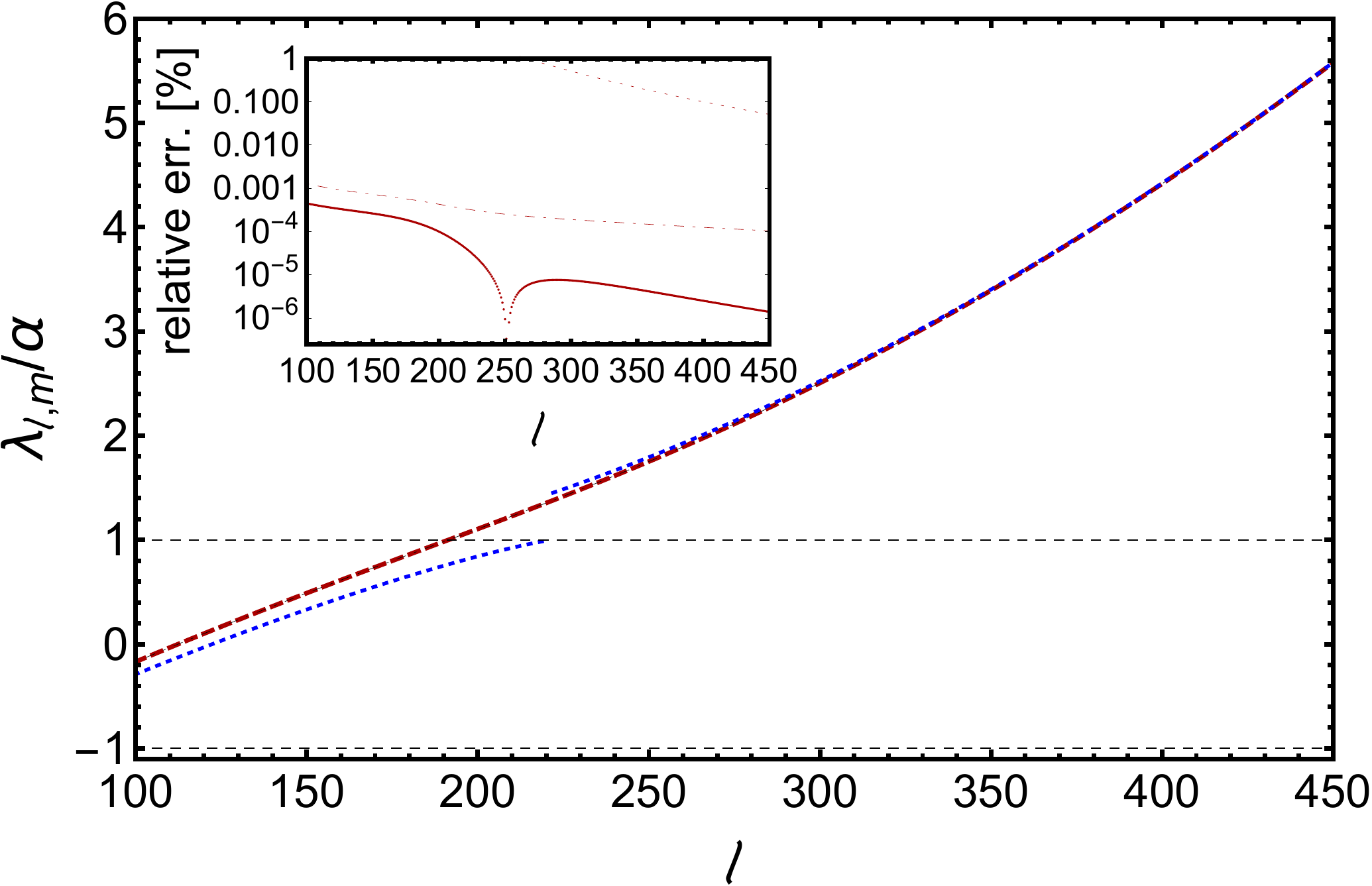}
	\caption{
		The spectrum $\lambda_{\ell,\,m}$ of $\hat U$ for $m=100$ in units of $\alpha$ calculated numerically for $\alpha = 3.65\times 10^4$ (dashed red line).
		Dot-dashed black line (coincides with full numerical dashed-red) is the spectrum obtained within quasi-classical approximation (see Eq.~\eqref{BS}).  Dotted blue
		lines represent solutions given by Eqs.~\eqref{BS_1} and~\eqref{BS_2}, below $\lambda<\alpha$ and above $\lambda>\alpha$,
		respectively. The inset shows the absolute relative error between solutions given by numerical diagonalization and: red dots -- quasi-classical approximation given by
		Eq.~\eqref{BS} (smallest relative error), red dot-dashed -- quasi-classical approximation with omitted shift 1/4, i.e., $\lambda = \varepsilon - \alpha$ (medium relative
		error), and red dotted line - the asymptotic form given by Eq.~\eqref{asymptot} (largest relative error).  
	}\label{fig:sp2}
\end{figure}

Figs.~\ref{fig:wf20} and \ref{fig:wf100} compare the wave functions $|\chi_{\ell,\, m}|^2$ obtained from Eqs.~\eqref{Ycl} and \eqref{chi}, with expansion coefficients
calculated numerically from Eq.~\eqref{eigenprob} and the quasi-classical wave functions Eq.~\eqref{CLacc}--\eqref{CLinaccL}. Both approaches agree in the whole region
except the neighbourhood of the classical turning points, where the quasi-classical approximation breaks down. The eigenvalues $\varepsilon_{\ell,m}$ for the
quasi-classical wave functions were obtained from Bohr-Sommerfeld's quantization rule, Eq.~\eqref{BS}, which works remarkably well, with relative errors smaller than
$10^{-3} \%$ (see Figs.~\ref{fig:sp1} and \ref{fig:sp2}).

Finally, we remark that the case $m=0$ should be treated with care. In the quasi-classical approximation, for $m=0$ the potential $v(\theta)$ has no classical turning points around $\theta =0$ and $\theta = \pi$. This can be traced back to dropping the contribution from $-\frac14 \sin^{-2}\theta$ term in the Schr\"odinger equation. In particular, the quasi-classical approximation that would be required to vanish for $\theta = 0$, where the potential $v(\theta)$ is finite, would have wrong phase in the classically allowed regime. Since for $\alpha \gg 1$ the contribution to collision rates comes typically from several partial waves with different values of $m$, here, we omit the quasi-classical analysis for $m=0$ and, when needed, refer to numerical calculations.

%
%
\begin{figure}[htb!]
	\includegraphics[clip, scale=0.55]{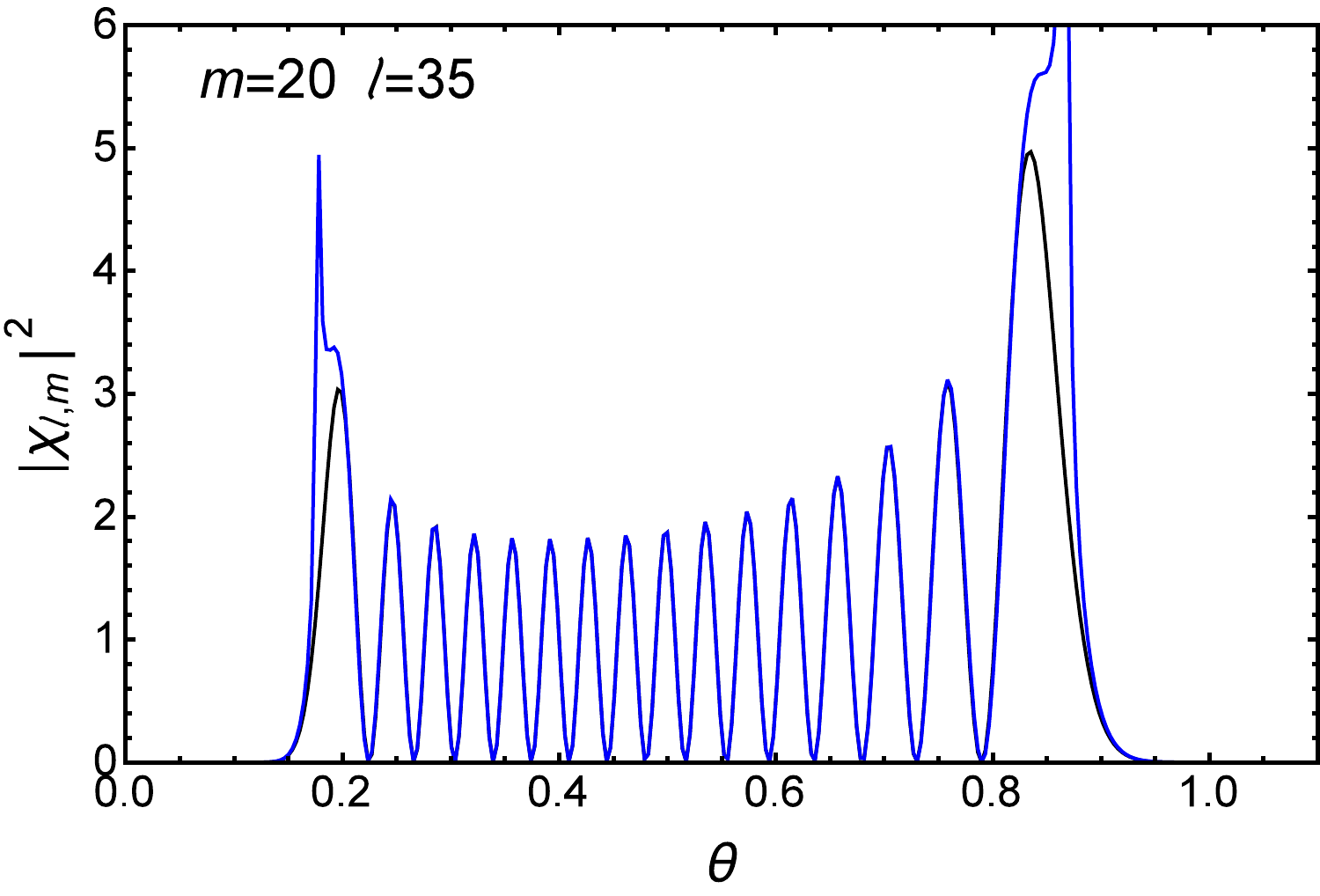}\\
	\includegraphics[clip, scale=0.55]{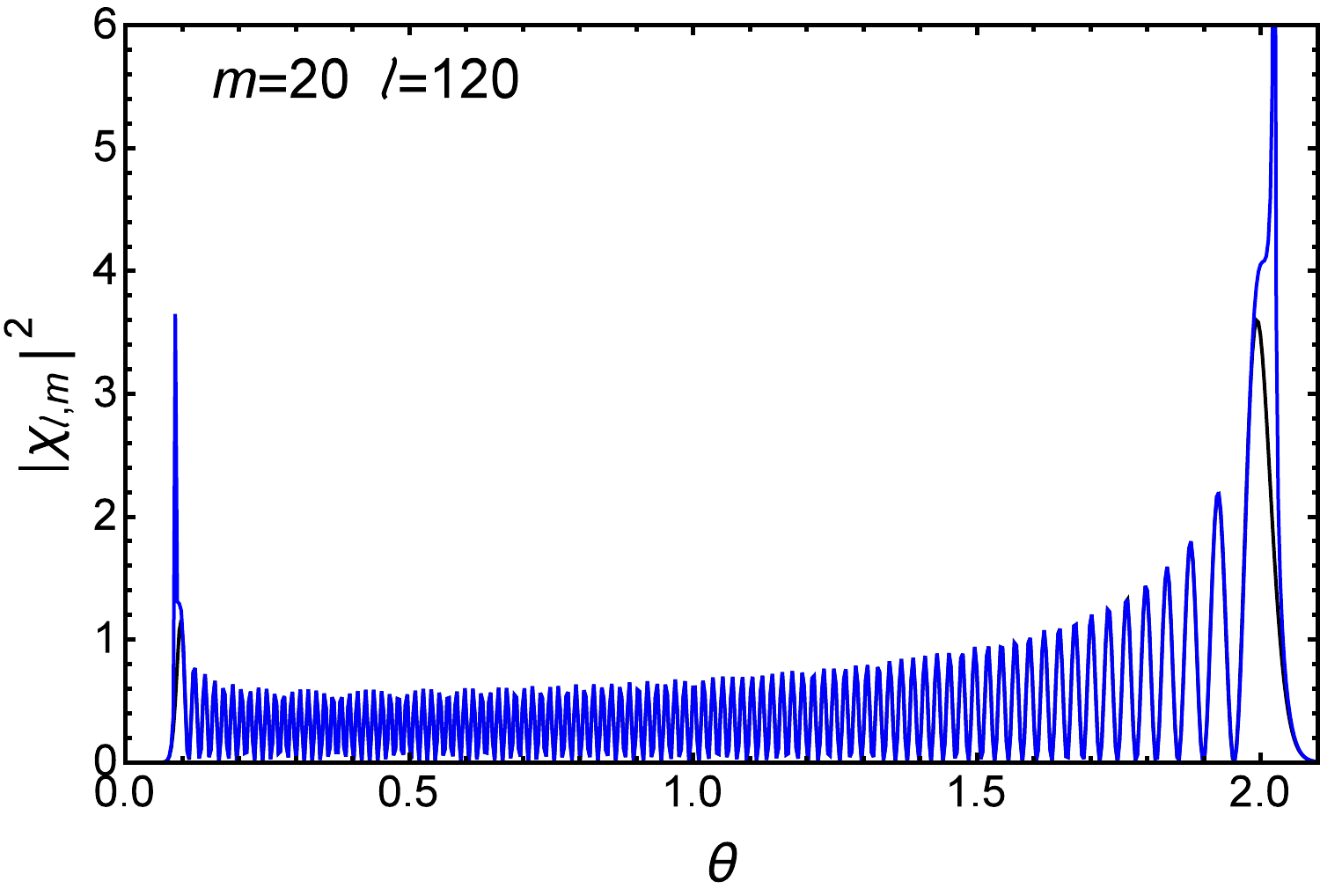}
	\caption{
		A comparison of the angular part of the wave function $|\chi_{\ell,\,m}|^2$ for $m=20$ and $\ell=35$ (upper panel) and $120$ (lower panel), which have $n=15$
		and $n=100$ nodes, respectively.  The black line is the solution numerically calculated from Eq.~\eqref{Ycl}, with the expansion coefficients $c_{\ell,\ell'}^{(m)}$ obtained by direct numerical diagonalization of \eqref{matU}. 
		The blue line is obtained within quasi-classical approximation using formulas~\eqref{CLacc}, \eqref{CLinaccR} and \eqref{CLinaccL}.  
	}\label{fig:wf20}
\end{figure}

%
%
\begin{figure}[htb!]
	\includegraphics[clip, scale=0.55]{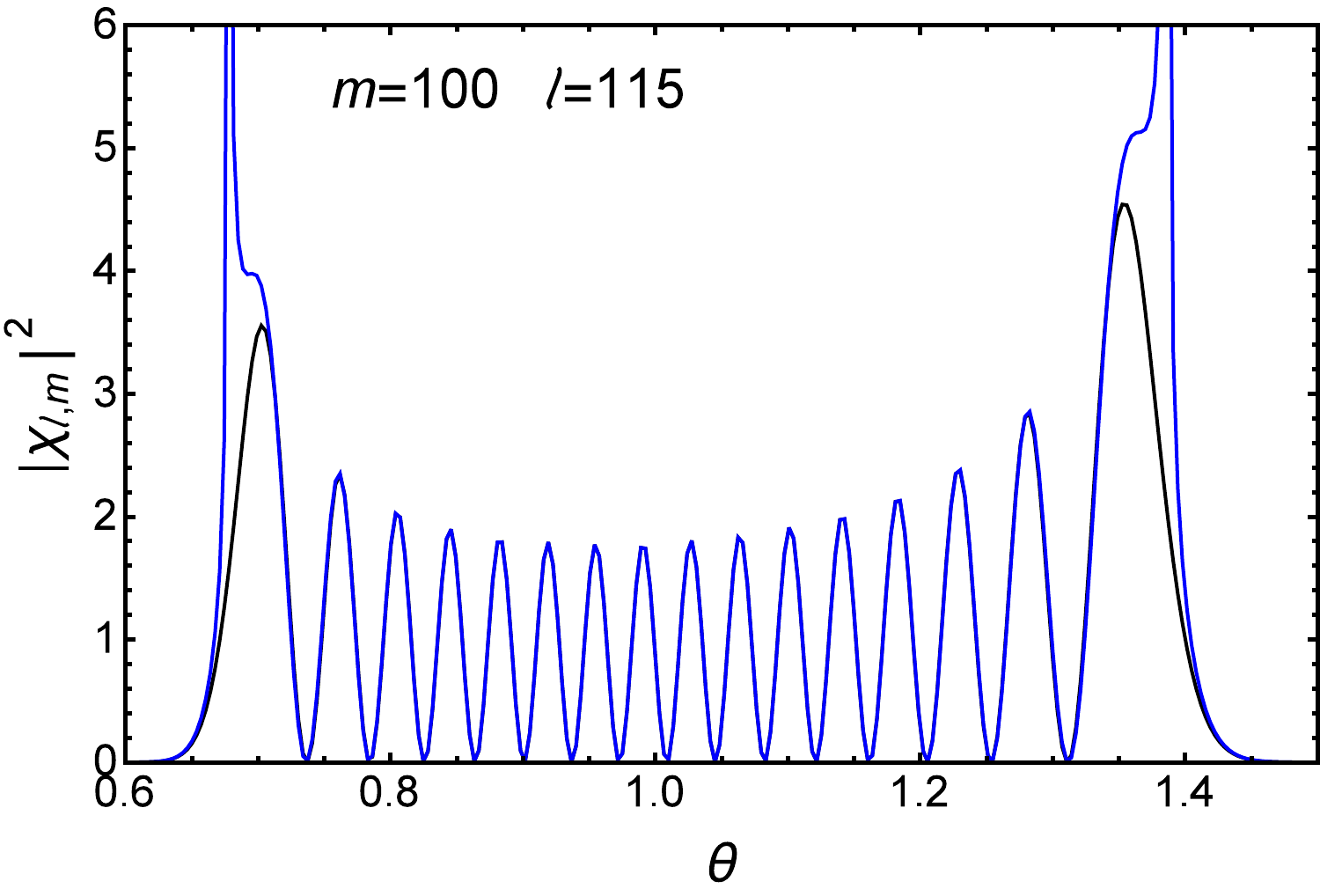}\
	\includegraphics[clip, scale=0.55]{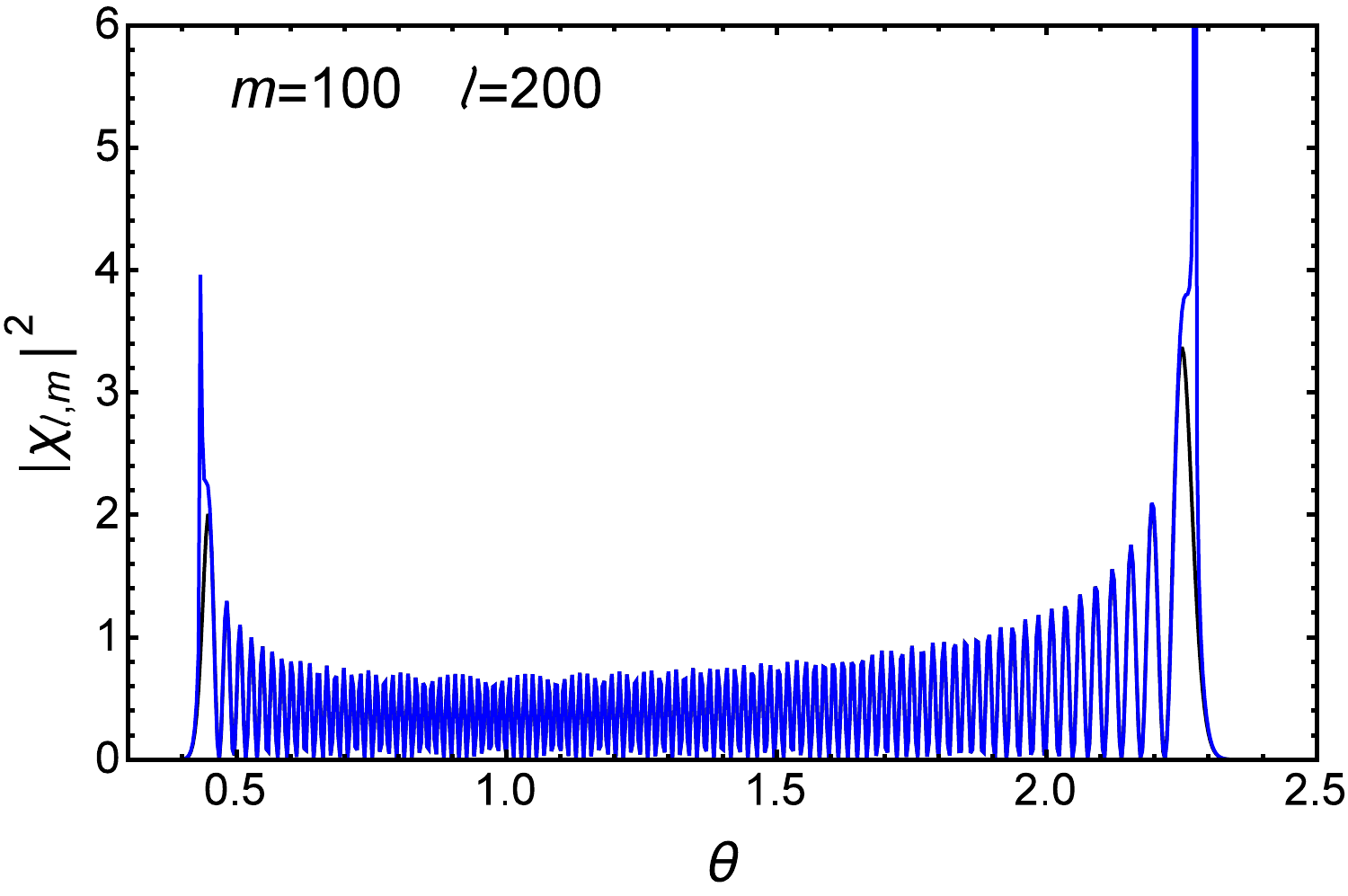}
	\caption{
		The comparison of the angular part of the wave function $|\chi_{\ell,\,m}|^2$ for $m=100$ and $\ell=115$ (upper panel) and $200$ (lower panel), which have
		$n=15$ and $n=100$ nodes, respectively. The black line is the solution numerically calculated from Eq.~\eqref{Ycl}, with the expansion coefficients $c_{\ell,\ell'}^{(m)}$ obtained by direct numerical diagonalization of \eqref{matU}. The blue line is obtained within quasi-classical approximation using formulas~\eqref{CLacc}, \eqref{CLinaccR} and \eqref{CLinaccL}.
	}\label{fig:wf100}
\end{figure}

}

%
%
\section{Derivation of continuous-$l$ approximation.}
\label{app-cont-l}
To understand the roles played by different terms in Eq.~\eqref{eigenprob}, we first discuss the behaviour of $c_{\ell,\,l}^{(m)}$ in the absence of the centrifugal
barrier $l(l+1)$ on the left hand side in Eq.~\eqref{eigenprob}. It will then turn out that $\hat l^2$ introduces an effective cutoff in the $l$ domain.  To start, we
write the equation for the eigenvalues without the term $\hat l^2$ in $\hat U$. The Eq.~\eqref{eigenprob} for large $l\gg 1$ then becomes
\begin{equation}
	- \frac{1}{2} \alpha\big( c_{\ell,\,l-1}^{(m)} + c_{\ell,\,l+1}^{(m)} \big) = \lambda_{\ell,\,m} c_{\ell,\,l}^{(m)},
\end{equation}
because $\beta_{m,l} \approx \frac12$ in this limit. This can be solved taking $c_{\ell,\,l}^{(m)} \sim e^{i l \tilde\theta}$ with $\lambda_{\ell,\, m}= - \alpha
\cos\tilde\theta$. The wave functions are then double degenerate in this limit because $c_l$ with the change $\tilde\theta \to - \tilde\theta$, have the same energy. Thus
two real solutions are $\sin l \tilde\theta$ and $\cos l \tilde\theta$. We reach the conclusion that in the absence of centrifugal barrier $|c_l|$ do not converge for
large $l$, which means that all partial waves contribute to the solution.

Another way to understand the effect is that the operator $\hat U$ contains $-\alpha\cos\theta$ only with $\hat l^2$ neglected. The solution of the eigenproblem $-\alpha\cos\theta \tilde Y(\theta,\phi) = \lambda \tilde Y(\theta,\phi)$ is given by
eigenvectors $\tilde Y(\theta,\phi) \propto \delta(\theta - \tilde\theta)$ with eigenvalues $\lambda= - \alpha \cos\tilde\theta$; the spectrum is continuously indexed with $\tilde\theta$. The lowest lying eigenvalue is $\lambda = -\alpha$ with $\tilde Y \sim \delta(\theta)$,
a function that is peaked around $\theta=0$.

The centrifugal barrier gives an effective cutoff for possible values of $l$ in $c_{\ell,\,l}^{(m)}$. Having this in mind, we now proceed to solve Eq.~\eqref{eigenprob} systematically in the limit of $l \gg q$. We start with the initial equation \eqref{eigenprob} with $\beta\approx 1/2$.
Assuming now that $c_{\ell,\,l}^{(m)}$ is a function $c(l)$ (we drop indices $\ell$ and $m$ and rename $\lambda_{\ell,\,m}$ to $\lambda$) we can expand $c(l\pm1)$ up to second order
for $l\gg|m|$. We therefore obtain the equation:
\begin{equation}
	- \frac{\alpha}{2} c''(l)  + \big(l(l+1) - \alpha \big) c(l)  = \lambda c(l).
\end{equation}
This is the Schr\"odinger equation for a shifted one-dimensional harmonic oscillator in which the position is given by $l$ and the wave function is $c(l)$. The solution
that converges for large $l$ is given by
\begin{equation}
	c(l) \propto D_\nu\!\bigg(  \frac{2l + 1}{(2 \alpha)^{1/4}} \bigg),
\end{equation}
where $D_\nu(x)$ is the parabolic cylinder function with $\nu = -1/2 + (1 + 4 \alpha + 4\lambda)/(4 \sqrt{2} \sqrt{\alpha})$.
The value of $\lambda$ and thus of $\nu$ is determined by the boundary conditions for $c(l)$ for small values of $l$ where $\beta_{m,l}$ deviates from $1/2$.
For large $x$, $D_\nu(x)$ falls off exponentially, so $c(l)$ becomes negligible for $l \gg \sqrt{\alpha}$. Also, from the obtained solution it is clear that $c(l)$ changes smoothly when incrementing $l$ by a unit.
These observations are the starting point for the following discussion.

Below, we expand the equation~\eqref{eigenprob} in the small parameter $\epsilon = \alpha^{-1/4}$.
\revM{As in the Appendix~\ref{app-cont-l-intro}}, we drop here the indices $\ell$ and $m$, rename $\lambda_{\ell,\,m}$ to $\lambda$ and write $c_{\ell,\,l}^{(m)}$ as $c(l)$.
We change the variables $x = \epsilon l$ and $c(l) = \tilde c(\epsilon l) \equiv \tilde c(x)$. We notice that for lowest levels $\lambda \approx -\alpha = -1/\epsilon^4$. Thus we write
$\lambda = \tilde \lambda/\epsilon^4$, where $\tilde \lambda$ is of the order of unity.
Now, we rewrite \eqref{eigenprob} in the following form:
\begin{widetext}
\begin{equation}
	- \frac{1}{\epsilon^2} \Bigg( \sqrt{ \frac{x^2 - \epsilon^2 m^2}{4 x^2 - \epsilon^2} } \tilde c(x-\epsilon) +  \sqrt{ \frac{(x+\epsilon)^2 - \epsilon^2m^2}{(2x+\epsilon)(2x+3\epsilon)} } \tilde c(x+\epsilon)  \Bigg)
	+ x(x+\epsilon)  \tilde c(x) = \frac{\tilde\lambda}{\epsilon^2} \tilde c(x).
\end{equation}
\end{widetext}
We expand the equation in the powers of $\epsilon$, which leads to Eq.~\eqref{eq_pos}.

%
%
\section{Quasi-classical analysis of the angular part}
\label{app-qcl}

In this appendix we derive formulas for the eigenvalues that are obtained from the Bohr-Sommerfeld's quantization rule, see Eq.~\eqref{BS}, for small $|m|$. 
We show that for small $\ell$ we recover the spectrum given by Eq.~\eqref{evals}. We also derive  a closed formula for large $\ell \gg \sqrt{\alpha}$ that
is independent of $m$, Eq.~\eqref{asymptot}.

If $|m|$ is not too large, the point $\theta_0$ where $\alpha(1-\cos\theta)$ is equal to $m^2/\sin^2\theta$, is much less than 1. If this is the case, then approximately
$\theta_0 \approx (2m^2/\alpha)^{1/4} \ll 1$. The following discussion is valid if $m^2 \ll \alpha$.

We first assume that \revM{$0\leqslant \varepsilon < 2\alpha$}.
Here, the position of the right turning point is mainly determined by the $\alpha(1-\cos\theta)$ term in $v(\theta)$, whereas $m^2/\sin^2\theta$ term gives only a small correction which we neglect, i.e.,
$\theta_R = \arccos(1-\varepsilon/\alpha)$.
The position of the left turning point is mainly determined by the $m^2/\sin^2\theta$ term in $v(\theta)$. Approximating $\sin\theta\approx \theta$
we find $\theta_L \approx (m^2/\varepsilon)^{1/2}$. We see that if $\varepsilon \gg |m| \sqrt{\alpha/2}$, the left turning point is separated from the point where the two parts of the potential
are of the same order, i.e., $\theta_L \ll \theta_0$.
We can use this separation, to effectively evaluate the dependence on $m^2$ from the integral. To see this, we notice, that the Born-Sommerfeld's quantization rules given by equation~\eqref{BS} can be
rewritten in the following form:
\begin{equation}
\label{BS2}
	\int_{\theta_L}^{\theta_L'} \!\! k_\mathrm{cl}(\theta') d\theta' + \int_{\theta_L'}^{\theta_R} \!\! k_\mathrm{cl}(\theta') d\theta' = \bigg(n+\frac12\bigg)\pi.
\end{equation}
Because of the mentioned separation, we choose $\theta_L'$ such that $\theta_L \ll \theta_L' \ll \theta_0$.
In this regime in the first integral on the left-hand side we can neglect $\alpha(1-\cos\theta)$ and
approximate $m^2/\sin^2\theta$ with $m^2/\theta^2$. Assuming that $\theta_L'^2 \gg m^2/\varepsilon$, we can evaluate the first integral to $\theta_L'\sqrt\varepsilon - |m|\pi/2$.
Now, in the second integral we can neglect the part $m^2/\sin^2\theta$ of the potential, and
write $\int_{\theta_L'}^{\theta_R} \!\! k_\mathrm{cl} = \int_{0}^{\theta_R} \!\! k_\mathrm{cl} - \int_{0}^{\theta_L'} \!\! k_\mathrm{cl}$. Here, the second integral
cancels the $\theta_L'\sqrt\varepsilon$ term from the first integral in~\eqref{BS2}. Finally, for $0\leqslant \varepsilon < 2\epsilon$ we obtain
\begin{equation}
\label{BS_1A}
	\int_{0}^{\theta_R} \!\! \sqrt{\varepsilon - \alpha(1-\cos\theta')} d\theta' = \bigg(2n+|m|+1\bigg)\frac{\pi}{2},
\end{equation}
with $\theta_R = \arccos(1-\varepsilon/\alpha)$.

As an application of formula~\eqref{BS_1}, we evaluate the low-lying eigenvalues. For small $\theta'$, we expand $1-\cos\theta'\approx\frac12\theta'^2$ in the integrand. The right turning point
$\theta_R = \sqrt{2\varepsilon/\alpha}$. Evaluation of the integral~\eqref{BS_1} is now straightforward. As a result we obtain the spectrum (neglecting the constant shift $-1/4$) that is
the same as the one given in equation~\eqref{evals}.

For the eigenvalues $\varepsilon\geqslant2\alpha$ we have to take care of the right turning point appropriately.
Using the same reasoning to the right region that is accessible to the particle, as we described in the paragraph above, we arrive at the following formula:
\begin{equation}
\label{BS_2A}
	\int_{0}^{\pi}\!\!\! \!\! \sqrt{\varepsilon \!-\! \alpha(1\!-\!\cos\theta')} d\theta' \!=\! \bigg(\!n\!+\!|m|\!+\!\frac12\bigg)\pi \!=\! \bigg(\!\ell + \frac12\bigg)\pi.
\end{equation}
We see that the eigenvalues here depend only on the quantum number $\ell$. We emphasize that these results are correct, if $|m|$ is not too large, i.e., $m^2 \ll \alpha$.
Note, that in the special case $\alpha=0$, we obtain $\varepsilon = (\ell+1/2)^2$, which gives exactly the correct value $\ell(\ell+1)$ if we reintroduce the $-1/4$ shift
to the eigenvalues of the angular part.

Using formula~\eqref{BS_2} we find the asymptotic form for the eigenvalues for large $\ell$. We start by introducing a small parameter $\alpha/\varepsilon$.
Expansion of the integrand on the left hand side up to the second power in this parameter leads to
$\sqrt{\varepsilon}(1 - \frac12 \frac{\alpha}{\varepsilon} - \frac{3}{16} \frac{\alpha^2}{\varepsilon^2}) = (\ell+\frac12)$. Squaring and retaining on the left hand side only terms up to the second order in
$\alpha/\varepsilon$ leads to
\begin{equation}
\label{app-asymptot}
	\varepsilon_\ell = \alpha + \bigg(\ell+\frac12\bigg)^2 + \frac{\alpha^2}{8(\ell+1/2)^2}.
\end{equation}
This equation is valid if the parameter $\alpha/\varepsilon_\ell$ is small, which is the case for $\ell \gg \sqrt\alpha$.

%
%
\section{The integral $\int \!\tilde Y_{\ell,m}^*(-\mathbf{n}) \tilde Y_{\ell,m}(+\mathbf{n})d\Omega$}
\label{app-YY}

Here we find the scalar product between the angular wave functions $\tilde Y_{\ell,m}(-\mathbf{n})$ and $\tilde Y_{\ell,m}(\mathbf{n})$ within the quasi-classical approximation.
We denote the integral by
\begin{equation}
  I_{\ell,m} = \int \!\tilde Y_{\ell,m}^*(-\mathbf{n}) \tilde Y_{\ell,m}(+\mathbf{n})d\Omega.
\end{equation}
The modified spherical harmonics are given by
\begin{equation}
  \tilde Y_{\ell,m}(\mathbf{n}) \equiv \tilde Y_{\ell,m}(\theta,\phi) = \Theta_{\ell,\,m}(\theta) \frac{e^{i m \phi}}{\sqrt{2\pi}},
\end{equation}
where $\Theta_{\ell,\,m}(\theta)$ fulfils Eq.~\eqref{ThetaEQ}.
The reflection from $\mathbf{n}$ to $-\mathbf{n}$ corresponds to change $\theta \to \pi - \theta$ and $\phi \to \pi+\phi$, so that
\begin{equation}
  \tilde Y_{\ell,m}(-\mathbf{n}) \equiv \tilde Y_{\ell,m}(\pi - \theta,\pi+\phi) = (-1)^m\Theta_{\ell,\,m}(\pi-\theta) \frac{e^{i m \phi}}{\sqrt{2\pi}}.
\end{equation}
The integral is therefore
\begin{equation}
  I_{\ell,m} = (-1)^m \int_0^\pi\!\!d\theta \Theta_{\ell,\,m}(\theta) \Theta_{\ell,\,m}(\pi-\theta),
\end{equation}
where we used fact that $\Theta_{\ell,\,m}$ is real.

Now we make first approximation, that the integration spans over the region that is classically accessible, neglecting the regions where the function decays exponentially.
Therefore the left turning point is $\theta_L' = \max(\theta_L,\pi-\theta_R)$ and the right is $\theta_R' = \min(\theta_R,\pi-\theta_L)$.

We may note that always $\theta_L < \pi/2$, and so if $\theta_R<\pi/2$, the integral is negligible and we have $I=0$. At least for small values of $m$, the condition
$\theta_R =\pi/2$ is reached for $\varepsilon \approx \alpha$.  Therefore, the integral is significantly non-zero only when $\varepsilon>\alpha$. In this regime,
$\theta_L' = \pi-\theta_R$ and $\theta_R' = \theta_R$.

In the quasi-classical approximation, the integral is given by
\begin{equation}
  I = (-1)^m C^2 \int_{\theta_L'}^{\theta_R'}\!\! d\theta'\ \frac{ \cos[\Phi_1(\theta')] \,  \cos[\Phi_2(\theta')]  }{ \sqrt{ k_{\mathrm{cl}}(\theta')  k_{\mathrm{cl}}(\pi-\theta')  }},
\end{equation}
where the normalization constant $C^{-2} = \frac12 \int_{\theta_L}^{\theta_R}1/k_\mathrm{cl}(\theta')$. The phase is given by
\begin{equation}
  \Phi_1(\theta') = \int_{\theta'}^{\theta_R} \!\! k_\mathrm{cl}(\theta'') d\theta'' - \frac{\pi}{4},
\end{equation}
and $\Phi_2(\theta') = \Phi_1(\pi-\theta')$.

This oscillatory integral is of the form $\int d\theta' f(\theta')\cos[\Phi_1] \,  \cos[\Phi_2]$, where the phases $\Phi_1$ and $\Phi_2$ are large.
Here, $f(\theta') = (-1)^m C^2 \ [k_{\mathrm{cl}}(\theta')  k_{\mathrm{cl}}(\pi-\theta')]^{-1/2}$. We use the method of stationary point (steepest descend) to evaluate the integral.
The integral can be written as
\begin{equation}
  I = \frac14 \int d\theta f(\theta) \bigg( e^{i\Phi} + e^{-i\Phi} + e^{i(\Phi_1-\Phi_2)} + e^{-i(\Phi_1-\Phi_2)}\bigg),
\end{equation}
where $\Phi(\theta) = \Phi_1(\theta) + \Phi_2(\theta)$.
The last two terms do not contribute, and we will neglect them. The stationary points can be only found for the first two. In order to find them, we equate the derivative of the total phase
to zero, i.e.,
\begin{equation}
  \Phi'(\theta_0) = -k_\mathrm{cl}(\theta_0) + k_\mathrm{cl}(\pi - \theta_0) = 0.
\end{equation}
The stationary point is $\theta_0 = \pi/2$.
Around this point, we may expand
\begin{equation}
  \Phi(\theta) = 2\Phi_1\Big(\frac{\pi}{2}\Big)
  + \Phi_1''\Big(\frac{\pi}{2}\Big) \Big(\theta-\frac{\pi}{2}\Big)^2
  + \frac{1}{12} \Phi_1''''\Big(\frac{\pi}{2}\Big) \Big(\theta-\frac{\pi}{2}\Big)^4.
\end{equation}
The constant term is equal to
\begin{equation}
  2\Phi_1\Big(\frac{\pi}{2}\Big)  = 2 \int_{\pi/2}^{\theta_R} \!\! k_\mathrm{cl}(\theta'') d\theta'' - \frac{\pi}{2}.
\end{equation}
The second term contains
\begin{equation}
  \Phi_1''\Big(\frac{\pi}{2}\Big) = - k'_\mathrm{cl}(\pi/2) = \frac{ \alpha }{ 2 \sqrt{ \varepsilon - \alpha - m^2 } }.
\end{equation}
Note that the dependence on the quantum numbers $\ell$ and $m$ is also contained in $\varepsilon=\varepsilon_{\ell,\,m}$.

The third term contains
\begin{eqnarray}
  && \Phi_1''''\Big(\frac{\pi}{2}\Big) =
  - \frac{\alpha}{2(\varepsilon- \alpha - m^2)^{1/2}} \\
  && \quad\quad+ \frac{3\alpha m^2}{2(\varepsilon - \alpha - m^2)^{3/2}}
   + \frac{3\alpha^3}{8(\varepsilon - \alpha - m^2)^{5/2}}.
\end{eqnarray}

The integral $I_{\ell,m}$ in the approximation is given by
\begin{equation}
  I_{\ell,m} = \frac12 f(\theta_0) \mathrm{Re}\Big( \int d\theta   e^{i\Phi(\theta)} \Big),
\end{equation}
which gives
\begin{equation}
  I_{\ell,m} =  \frac{(-1)^m C^2}{2k_{\mathrm{cl}}(\pi/2)} \mathrm{Re}\Big( \int_{\theta_L'}^{\theta_R'} d\theta   e^{i\Phi(\theta)} \Big),
\end{equation}
with $k_{\mathrm{cl}}(\pi/2) = \sqrt{\varepsilon-\alpha - m^2}$.
Inserting here the expansion, we obtain:
\begin{eqnarray}
  && I =  \frac{(-1)^m C^2}{2k_{\mathrm{cl}}(\pi/2)} \mathrm{Re}\Big\{
  e^{2i\Phi_1(\pi/2)} \times \\
  && \times
  \int_{\theta_L'}^{\theta_R'} d\theta
  e^{-i k'_\mathrm{cl}(\pi/2) (\theta-\theta_0)^2
    + \frac{i}{12} \Phi_1''''(\theta_0) (\theta-\theta_0)^4}
  \Big\}, \quad\quad.
\end{eqnarray}
Now, we assume that we can neglect the fourth-order term in $\theta$ in the exponent, and extend the integral to infinity. Using the formula $\int dx e^{i a x^2} = \sqrt{\pi/a} e^{i \pi/4}$, valid for $a>0$,
we obtain the results
\begin{eqnarray}
  && \int \!\tilde Y_{\ell,m}^*(-\mathbf{n}) \tilde Y_{\ell,m}(+\mathbf{n})d\Omega = \sqrt{\frac{\pi}{2\alpha}} \bigg[ \frac12 \int_{\theta_L}^{\theta_R} \frac{d\theta'}{k_\mathrm{cl}(\theta')}  \bigg]^{-1} \!\!\!\!\!\times \quad\quad \nonumber\\
  && \ \times \frac{(-1)^m}{( \varepsilon_{\ell,m} \!-\! \alpha \!-\! m^2 )^{1/4}} \cos\!\bigg( 2 \int_{\pi/2}^{\theta_R} k_\mathrm{cl}(\theta')d\theta' \!-\! \frac\pi4  \bigg), \quad\quad
  \label{YY1}
\end{eqnarray}
for $\varepsilon_{\ell,\,m} > \alpha$; for $\varepsilon_{\ell,\,m}<\alpha$ the integral vanishes exponentially and we approximate it by zero. This formula is valid as
long as the integration region can be extended to the real domain, which happens provided \mbox{$(|k'_\mathrm{cl}(\pi/2)|)^{-1/2} \ll \pi/2 \approx 1$}.

It is also possible to find the asymptotic expansion for large $\ell$, that is for large $\varepsilon_{\ell,\,m}$, when it takes the limiting form given by Eq.~\eqref{asymptot}.
We start, by writing the integral $I_{\ell,m} $ in the form:
\begin{equation}
  I_{\ell,m}  = (-1)^m \bigg\{ \frac{C^2 \pi}{ 2 \sqrt{\varepsilon - \alpha - m^2}} \bigg\} \times \mathrm{Re}\bigg[ e^{2 i \Phi_1(\pi/2)} \times \mathcal{M} \bigg],
\end{equation}
where
\begin{equation}
  \mathcal{M} = \frac{1}{\pi} \int_{-\pi/2}^{+\pi/2}\!\! e^{i a x^2 + i b x^4}\, dx,
\end{equation}
with $a = \alpha / [2(\varepsilon - \alpha - m^2)^{1/2}]$ and $b \approx -  \alpha / [ 24 (\varepsilon-\alpha-m^2)^{1/2} ] $.
The factor in curly brackets rapidly approaches 1, as $\epsilon \to \infty$, and consequently, we drop it.
For large $\ell$ we may approximate $\varepsilon -\alpha - m^2 \approx (\ell+1/2)^2$ in the following discussion. The small parameter in the expansion is set then by $\epsilon = \alpha/(\ell+1/2)$.
Using the same line of reasoning, which led to Eq.~\eqref{asymptot}, we evaluate
\begin{equation}
  2  \Phi_1(\pi/2) = \pi(\ell - |m|) - \epsilon.
\end{equation}
The parameter $a$ and $b$ are given by $a\approx \epsilon/2$ and $b \approx -\epsilon/24$. Now, we expand the exponent under the integral in $\mathcal{M}$ up to the second order in $\epsilon$ and expand the exponential term 
$e^{2 i \Phi_1(\pi/2)} $ Then, collecting terms up to the second power of $\epsilon$, and, finally, taking the real part, yields
\begin{equation}
\label{YYasymptot}
I_{\ell,m}  \xrightarrow[\ell \to \infty]{}  (-1)^\ell \bigg[  1 - \zeta_0 \bigg( \frac{\alpha}{\ell+1/2} \bigg)^2\bigg],
\end{equation}
where the number $\zeta_0 = 1/2 - \pi^2/24 + \pi^4/480 - \pi^6/21504 + \pi^8/2654208 \approx 0.25057$. We note that in the limit of large $\ell$, integral $I_{\ell,m}$ tends to the value $(-1)^\ell$, identical as for standard spherical harmonics: $\int
Y_{\ell,m}^*(-\hat{\mathbf{k}}_i) Y_{\ell,m}(\hat{\mathbf{k}}_i) d\Omega_i = (-1)^\ell$.

\begin{figure}[htb!]
 \includegraphics[clip, scale=0.5]{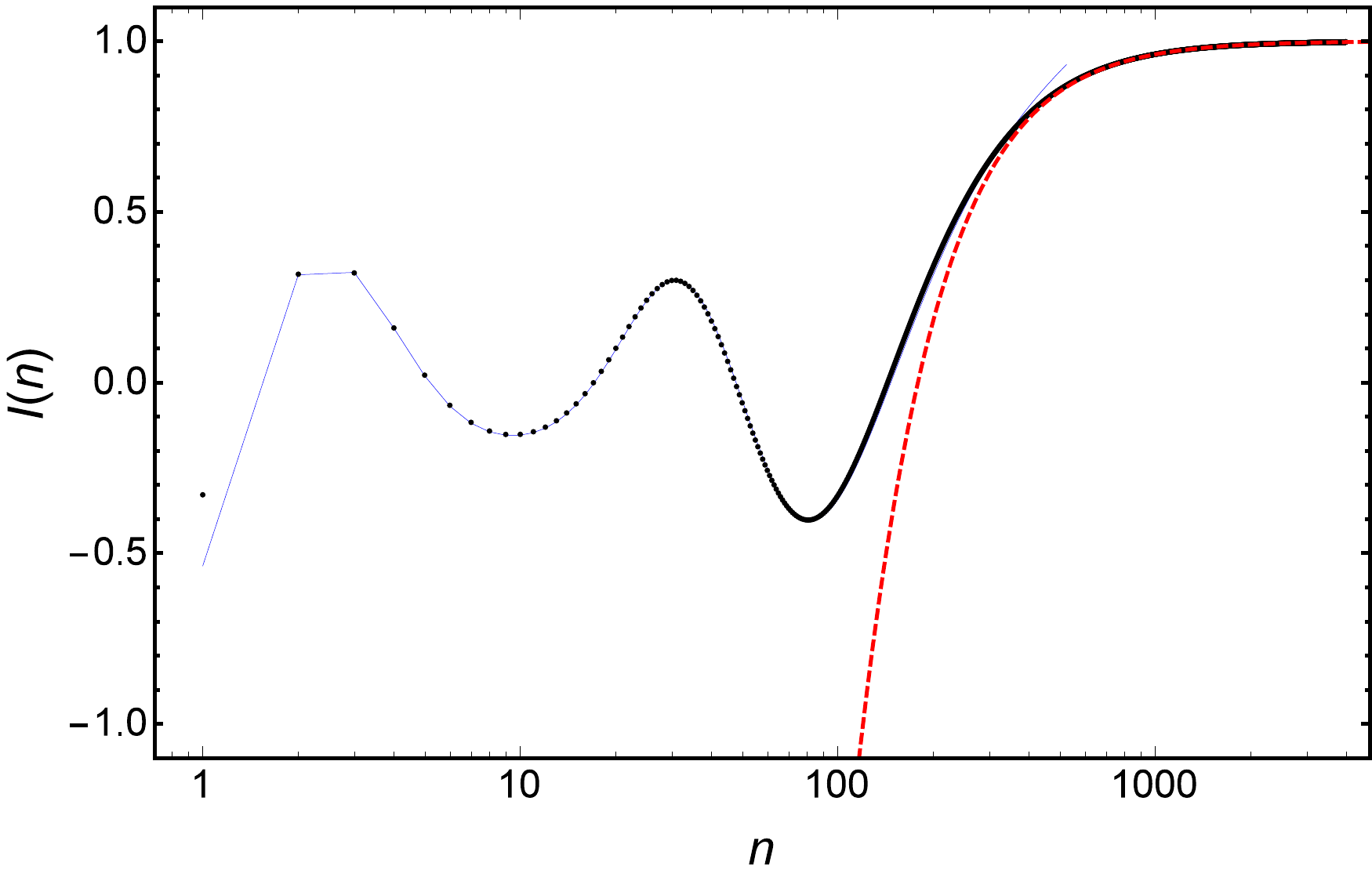}
  \caption{
    The scalar product $(-1)^\ell I_{\ell,m} = (-1)^\ell \int \!\tilde Y_{\ell,m}^*(-\mathbf{n}) \tilde Y_{\ell,m}(+\mathbf{n})d\Omega$ as a function of the quantum number $n=\ell-|m|=0,1,2,\ldots$.
    In this example $\alpha = 400$, $m = 21$.
    The black dots represents full numerical calculation, blue solid line -- quasi-classical approximation given by Eq.~\eqref{YY1}, and red dashed line -- asymptotic expansion given by Eq.~\eqref{YYasymptot}.
    }\label{fig:I}
\end{figure}

Fig.~\ref{fig:I} shows the values of integral $I_{\ell,m}$ multiplied by the factor $(-1)^\ell$ for simplicity. It compares exact values calculated numerically, with quasi-classical formula \eqref{YY1} and with large-$\ell$ expansion \eqref{YYasymptot}. We observe that quasi-classical formula \eqref{YY1} is very accurate up to $ n = \ell - |m| \lesssim \alpha$, when it starts to deviate from the exact values. In that regime, however, the asymptotic expansion \eqref{YYasymptot} can be already used.  
From this last result we infer, that $\bar\sigma_\mathrm{el} = \infty$. This is the consequence of the slow decay of $I_{\ell,m}$ with $\ell$, which is 
compensated by the sums over $m$ and $\ell$ in the elastic collision rate $K^\mathrm{el}$.

%
%

%
%
\section{Radial part: quasi-classical approximation.}
The radial Schr\"odinger equation can be solved within the quasi-classical approximation. We define the local wave vector as
\begin{equation}
	q_\mathrm{cl}(r) = \sqrt{k^2 - \frac{\varepsilon-\alpha}{r^2}},
\end{equation}
where $\varepsilon$ is one of the eigenvalues $\varepsilon_{\ell,\,m}$. Here, we neglected the $-\frac14$ term in $\lambda = \varepsilon-\alpha - \frac14$, as is usual in quasi-classical approximation.
Now, the radial solutions can be written as a superposition of the wave functions:
\begin{equation}
	R^{\pm}_\mathrm{cl}(r) = \frac{ e^{\pm i \big[\int  q_\mathrm{cl}(r)dr + |\kappa|\log[|\kappa|k]-\frac{\pi}{4} \big] } e^{\frac{|\kappa|\pi}{2}}  }{2r \sqrt{ k q_\mathrm{cl}(r) }}.
\end{equation}

We will now investigate the behaviour of this solution for large and small distances. We start with the case $\varepsilon<\alpha$ (attractive potential), then we can write \mbox{$\varepsilon-\alpha = - |\kappa|^2$}.
The integral in the exponents can be evaluated and reads
\begin{eqnarray}
	&&\int\!\!\sqrt{ k^2 + \frac{|\kappa|^2}{r^2} } dr
		= \sqrt{ (kr)^2 + |\kappa|^2 } + \\
		&& \quad\quad\quad+\ |\kappa| \log\!\bigg( \frac{r}{|\kappa|(|\kappa| + \sqrt{ (kr)^2 + |\kappa|^2 })} \bigg),
\end{eqnarray}
where the constant of integration was omitted, because it can be incorporated into normalization. For small values of $r \ll |\kappa|/k$, the $r$-dependent term in the integral
is $|\kappa| \log(kr)$. Therefore, the quasi-classical solution $R^{\pm}_\mathrm{cl}(r)$  corresponds~\footnote{To show that normalization constant is correct it is useful to exploit the asymptotic Stirling relation
$\Gamma(1+i x) \approx e^{i\pi/4} e^{-\frac{x}{2}(2 i + \pi)} \sqrt{2\pi} x^{\frac12 + i x}$, which is valid for $x\gg1$} to exact solution $R^{\pm}(r)$.
To see that the correspondence holds also for large $r$, it is easy to verify that the quasi-classical approximation holds if $|\kappa|^2/[(kr)^2 + |\kappa|^2] \ll 1$.
The quasi-classical approximation is always valid for large $r \gg |\kappa|^{2/3} / k$. However, if the approximation is valid for small $r$, which is equivalent to
$|\kappa| \gg 1$, due to inequality $|\kappa|^2/[(kr)^2 + |\kappa|^2] \leqslant 1 / |\kappa|$, it is also valid at large distances. In the latter case, the parameter $|\kappa| \gg 1$,
and for large distances, the term $e^{\mp ikr}$ is exponentially smaller than the term $e^{\pm ikr}$ in the exact~$R^{\pm}$.

Now we will consider the second branch of the angular spectrum, i.e. $\varepsilon\geqslant\alpha$, for which $\varepsilon-\alpha = |\kappa|^2$. In this regime,
we have a classically accessible and inaccessible regions separated by the classical turning point, ${r_\mathrm{tp} = |\kappa|/k}$.
The quasi-classical wave function $R^{+}_\mathrm{cl}(r)$, away from the turning point, now reads
\begin{equation}
	R^{+}_\mathrm{cl}(r) = \frac{ e^{- \int_{r}^{r_\mathrm{tp}}  |q_\mathrm{cl}(r')|dr'} }{2 r \sqrt{ k |q_\mathrm{cl}(r)| }},
\end{equation}
for $r < r_\mathrm{tp}$, and
\begin{equation}
	R^{+}_\mathrm{cl}(r) = \frac{ \cos\!\big( \int_{r_\mathrm{tp}}^{r}  q_\mathrm{cl}(r')dr' - \frac{\pi}{4} \big)}{r \sqrt{ k q_\mathrm{cl}(r) }},
\end{equation}
for $r>r_\mathrm{tp}$.
The wave function $R^{+}_\mathrm{cl}(r)$ corresponds to exact solution ~$R^{+}_{k,\ell,m}(r)$.
The second solution $R^{-}_\mathrm{cl}(r)$ for $r>r_\mathrm{cl}$ is
\begin{equation}
	R^{-}_\mathrm{cl}(r) = \frac{ \cos\!\big( \int_{r_\mathrm{tp}}^{r}  q_\mathrm{cl}(r')dr' - \frac{\pi}{4}  + \frac{|\kappa|\pi}{2} \big)}{r \sqrt{ k q_\mathrm{cl}(r) }},
\end{equation}
while for $r<r_\mathrm{cl}$ is
\begin{equation}
	R^{-}_\mathrm{cl}(r) = \frac{ e^{+ \int_{r}^{r_\mathrm{tp}}  |q_\mathrm{cl}(r')|dr'} }{ r \sqrt{ k |q_\mathrm{cl}(r)| }} \mathcal{A}^{(-)},
\end{equation}
where the amplitude, which matches the wave function in the classically allowed region, is given by
\begin{equation}
	\mathcal{A}^{(-)} =  \sqrt{\frac{\pi}{4}}\frac{1}{\Gamma(1-|\kappa|)} \bigg(\frac{|\kappa|}{e}\bigg)^{-|\kappa|}2^{1/2}\sqrt{|\kappa|}.
\end{equation}
This wave function corresponds to $R^{-}_{k,\ell,m}(r)$.



%

\end{document}